\documentclass[11pt]{extarticle}
\usepackage[left=2cm, right=2cm, top=2cm]{geometry}

\usepackage[pdftex]{graphicx}
\graphicspath{ {./images/} }

\usepackage{lineno}
\usepackage{floatrow}
\usepackage{multirow}
\usepackage{multicol}
\usepackage{adjustbox}
\usepackage{lipsum}
\usepackage{siunitx}
\usepackage{caption}

\usepackage{subcaption}
\usepackage[fleqn]{amsmath}
\usepackage{setspace}
\usepackage[version=4]{mhchem}

\usepackage{rotating}
\usepackage{csquotes}
\usepackage{chemformula}
\usepackage{longtable}
\usepackage{pdflscape}

\usepackage{pifont}% http://ctan.org/pkg/pifont
\usepackage[nonumberlist]{glossaries}

\newacronym{AoC}{AoC}{Areas of Concern}
\newacronym{AoP}{AoP}{Areas of Protection}
\newacronym{CAT}{CAT}{Climate Action Tracker}
\newacronym{CC}{CC}{Climate Change}
\newacronym{CF}{CF}{Carbon Footprint}
\newacronym{COP21}{COP21}{Conference of the Parties 2021}
\newacronym{EQ}{EQ}{Ecosystem Quality}
\newacronym{ES}{ES}{EnergyScope}
\newacronym{EP50+}{EP50+}{Energieperspektiven 2050 +}
\newacronym{FNEU}{FNEU}{Fossil and Nuclear Energy Use}
\newacronym{FU}{FU}{Functional Unit}
\newacronym{GHG}{GHG}{Greenhouse Gas}
\newacronym{GWP}{GWP}{Global Warming Potential}
\newacronym{HH}{HH}{Human Health}
\newacronym{IPCC}{IPCC}{Intergovernmental Panel on Climate Change}
\newacronym{IRENA}{IRENA}{International Renewable Energy Agency}
\newacronym{IW+}{IW+}{IMPACT World+}
\newacronym{LCA}{LCA}{Life Cycle Assessment}
\newacronym{LCI}{LCI}{Life Cycle Inventory}
\newacronym{LCIA}{LCIA}{Life Cycle Impact Assessment}
\newacronym{MILP}{MILP}{Mixed Integer Linear Programming}
\newacronym{MOO}{MOO}{Multi-Objective Optimization}
\newacronym{NDCs}{NDCs}{Nationally Determined Contributions}
\newacronym{OF}{OF}{Objective Function}
\newacronym{PV}{PV}{Photovoltaic Panel}
\newacronym{REQD}{REQD}{Remaining Ecosystem Quality Dammage}
\newacronym{RHHD}{RHHD}{Remaining Human Health Dammage}
\newacronym{SOO}{SOO}{Single-Objective Optimization}
\newacronym{WSF}{WSF}{Water Scarcity Footprint}

\makeglossaries

\makeglossaries
\usepackage[flushleft]{threeparttable}

\usepackage{array}
\newcolumntype{L}[1]{>{\raggedright\let\newline\\\arraybackslash\hspace{0pt}}m{#1}}
\newcolumntype{C}[1]{>{\centering\let\newline\\\arraybackslash\hspace{0pt}}m{#1}}
\newcolumntype{R}[1]{>{\raggedleft\let\newline\\\arraybackslash\hspace{0pt}}m{#1}}

\usepackage{textcomp}

\usepackage{amssymb}

\usepackage{hyperref}
\hypersetup{
    colorlinks=true,
    linkcolor=blue,
    filecolor=magenta,      
    urlcolor=blue,
}

\usepackage[british]{babel}

\usepackage{amsmath}
\usepackage{breqn}\usepackage{booktabs}

\usepackage{enumitem,kantlipsum}
\usepackage{algorithm}
\usepackage{algpseudocode}
\usepackage[backend=biber,natbib=true,style=numeric-comp,citestyle=numeric-comp,terseinits=true,uniquename=init,giveninits=true,sorting=none,doi=false,isbn=false,url=false,maxbibnames=15]{biblatex}
\bibliography{main}
\usepackage{colortbl}
\usepackage{tcolorbox} % to create boxes in the text\usepackage{comment}
\definecolor{rouge}{HTML}{FF0000}
\definecolor{groseille}{HTML}{B51F1F}
\definecolor{canard}{HTML}{007480}
\title{Between Green Hills and Green Bills: Unveiling the Green Shades of Sustainability and Burden Shifting through Multi-Objective Optimization in Swiss Energy System Planning}

\author{Jonas Schnidrig$^{1,2^{*,\ddagger}}$, Matthieu Souttre$^{3,\ddagger}$, Arthur Chuat$^{1,2,\ddagger}$\\François Maréchal$^{1,3}$ and Manuele Margni$^{2,3}$ }
\date{1st December 2023}
\begin{document} 

\maketitle 

$^1$Industrial Process and Energy Systems Engineering group, École Polytechnique 
Fédérale de Lausanne, Rue de l'Industrie 17, 1950 Sion, Switzerland;

$^2$CIRAIG, Institute for Sustainable Energy, University of Applied Sciences Western Switzerland, Rue de l'Industrie 23, 1950 Sion, Switzerland;

$^3$CIRAIG, École Polytechnique de Montreal, 3333 Queen Mary Rd, Montréal, Québec;

$^*$jonas.schnidrig@hevs.ch\\
$^\ddagger$ Authors contributed equally

\vspace{1cm}

\begin{abstract}
\noindent
The Paris agreement is the first-ever universally accepted and legally binding agreement on global climate change. It is a bridge between today's and climate-neutrality policies and strategies before the end of the century. Critical to this endeavor is energy system modeling, which, while adept at devising cost-effective carbon-neutral strategies, often overlooks the broader environmental and social implications. This study introduces an innovative methodology that integrates life-cycle impact assessment indicators into energy system modeling, enabling a comprehensive assessment of both economic and environmental outcomes.

\noindent
Focusing on Switzerland's energy system as a case study, our model reveals that optimizing key environomic indicators can lead to significant economic advantages, with system costs potentially decreasing by \SI{15}{\%} to \SI{47}{\%} by minimizing potential impacts from operating fossil technologies to the indirect impact related to the construction of the renewable infrastructure. However, a system optimized solely for economic efficiency, despite achieving \SI{63}{\%} reduction in carbon footprint compared to 2020, our results show a potential risk of burden shift to other environmental issues.

\noindent
The adoption of multi-objective optimization in our approach nuances the exploration of the complex interplay between environomic objectives and technological choices. Our results illuminate pathways towards more holistically optimized energy systems, effectively addressing trade-offs across environmental problems and enhancing societal acceptance of the solutions to this century's defining challenge.

\end{abstract}

\vspace{1cm}

%%%%%%%%%%%%%%%%%%%%%%%%%%%%%%%%%%%%%%%%%%%%%%%%%%%%%%%%%%%%%%%%%%%%%%%%%%%%%%
%%%%%%%%%%%%%%%%         RESULTS AND FINDINGS        %%%%%%%%%%%%%%%%%%%%%%%%%
%%%%%%%%%%%%%%%%%%%%%%%%%%%%%%%%%%%%%%%%%%%%%%%%%%%%%%%%%%%%%%%%%%%%%%%%%%%%%%

\section{Introduction}

% Background
\subsection{Background}
The escalation in the intensity and frequency of climate change events underscores the critical need for robust mitigation strategies. The \gls{IPCC} emphasizes a global effort to curtail \gls{GHG} emissions, aiming to contain global warming within 1.5 - 2°C above pre-industrial levels. The Paris Agreement, formulated at COP21 in 2015, mandates signatories to present increasingly ambitious \gls{NDCs} every five years to articulate mid to long-term emissions reduction goals. Institutions such as \gls{CAT} and the \gls{IRENA} assess the alignment of \gls{NDCs} with the targeted temperature thresholds and offer guidance for enhancements. With a growing focus on sustainable development and the shift toward renewable energy sources, energy system modeling has garnered significant interest. Concurrently, the \gls{LCA} approach, which evaluates potential impacts of products and technologies over the whole value chain on a broader spectrum of environmental impacts like ozone depletion and particulate matter formation, has grown substantially over the past two decades.

To genuinely transition to a low-carbon energy system, accounting for and monitoring potential environmental burden shifts beyond \gls{GHG} emissions is imperative. Our methodology seeks to meld \gls{LCA} within Energy System Modeling, optimizing the energy system's carbon footprint, water scarcity footprint, fossil \& nuclear energy use, and remaining damages to human health and ecosystem quality. 

%literature review 
\subsection{Literature Review}
% From small to large scale systems
Energy system design traditionally pivots around economic optimization. However, in the wake of growing environmental concerns and sustainability priorities, there has been a discernible shift in focus towards more comprehensive assessment methods. \gls{LCA}, which primarily found applications in analyzing small-scale technologies \cite{laurentLCAEnergySystems2018,petrilloLifeCycleAssessment2016,valenteLifeCycleAssessment2017}, is rapidly gaining prominence in evaluating larger systems such as processes \cite{pieragostiniProcessOptimizationConsidering2012}, plants \cite{wangLowCarbonOptimal2022}, and buildings \cite{mayerEnvironmentalEconomicMultiobjective2020}.

% History of ESys-LCA coupling
\gls{LCA} has been comprehensively coupled with an energy system model for the first time by Loulou et al. \cite{loulou2008} within the NEEDS project in 2008, in order to assess the external costs of power production. Since then, numerous energy system models have been coupled to \gls{LCA}, with varying integration levels, environmental comprehensiveness and sectoral coverage. Most energy system models are soft-linked to \gls{LCA}, i.e., the output of the model feed the \gls{LCA} model, resulting in an environmental profile of the energy system, which may eventually be given back to the energy model until a convergence criterion is met. However, a unique iteration is typically performed, thus resulting in a so-called ex-post analysis. Ex-post analyses \cite{baumgartner2021,huber2023,naegler2022,blanco2020,fernandezastudillo2019,boubault2019,volkart2018,garcia-gusano2017,pehl2017,berrill2016} are static by nature and thus limited regarding the use of \gls{LCA} metrics in the energy model. A few models are hard-linked to \gls{LCA}, i.e., \gls{LCA} metrics are endogenously integrated into the model, thus allowing to use them in the objective function or constraints in the case of an optimization model. \gls{LCA} metrics can be integrated within the optimization problem using three main techniques: 1) the normalization and weighting of different objectives in the objective function \cite{vandepaer2020,louis2020,rauner2017,barteczko-hibbert2014} ; 2) the $\epsilon$-constraint method \cite{vandepaer2020,tokimatsu2020}, i.e., optimize for a single objective and set an upper bound to other objectives as constraints ; 3) the monetization of environmental objectives \cite{algunaibet2019b}. On one hand, Vandepaer et al. \cite{vandepaer2020} optimized the sum of two objectives, namely the system total cost and the so-called total cumulative LCIA score. A main limitation is the cost weighting factor that had to be arbitrarily chosen to allow the model convergence. Vandepaer et al. further used the $\epsilon$-constraint method by setting upper bounds to objectives, which were chosen using the minimum value derived from the single-objective optimizations of those, and adding a relaxation term varying between 5 and 50\%. On the other hand, Algunaibet et al. \cite{algunaibet2019b} monetized three damage-level indicators, namely human health, ecosystem quality and resource availability, to include them in an economic objective function. Even if monetization avoids the normalization and weighting steps, the conversion of environmental impact into monetary terms remains highly uncertain. 

The coupling between energy models and \gls{LCA} being complex, most analyses started by focusing on the electricity sector alone \cite{barteczko-hibbert2014,shmelev2016,portugal-pereira2016,berrill2016,garcia-gusano2016,pehl2017,rauner2017,mcdowall2018,louis2018}. However, a few studies now include multi-energy models \cite{volkart2018,fernandezastudillo2019,blanco2020,vandepaer2020,naegler2022,terlouw2023,reinert2023}, thus accounting for the inter-dependencies of the energy sectors, namely electricity, heat and mobility. Still, using a too wide model may eventually result in an incomplete mapping of the model's technologies with life-cycle inventories. This issue leads to only using a sub-set of most contributing processes \cite{fernandezastudillo2019} or having non-characterized, and thus environmentally favored, energy technologies \cite{vandepaer2020,louis2018}.

The comprehensiveness of environmental impact categories varies significantly across studies and strongly depends on their research questions and objectives. While some solely focus of \gls{GHG} emissions \cite{portugal-pereira2016,menten2015}, other use a full \cite{huber2023,blanco2020,louis2018,volkart2018,rauner2017,garcia-gusano2016a,algunaibet2019b} or selection \cite{martin2023,garcia-gusano2017,berrill2016} of midpoint-level indicators \cite{martin2023,huber2023,blanco2020,louis2018,volkart2018,rauner2017,berrill2016}, damage-level indicators \cite{garcia-gusano2016a,algunaibet2019b}, or a combination of those \cite{garcia-gusano2017}.  

However, a critical observation emerges from the current literature: while he coupling between \gls{LCA} and optimization tools is an active research area, none of the previously studied literature encompasses (i) a national model addressing all energy demands, (ii) the systematic and consistent integration of a comprehensive set of \gls{LCA} metrics, and (iii) consideration of uncertainties, which remains elusive.

\section{Problem Statement}

In the evolving landscape of energy system design, \gls{LCA} has become increasingly critical, particularly in the transition from smaller-scale technologies to more comprehensive systems like processes, plants, and buildings. Traditional models often focus on economic optimization, with \gls{LCA} considerations taking a secondary role, especially in larger systems where economic objectives have predominated. This prioritization has led to a notable gap in the literature, where the integration of comprehensive environmental indicators with economic goals in energy system modeling, particularly in optimizing for these environmental indicators while addressing uncertainties, is still largely unexplored.

This project utilizes the energy system model \gls{ES}, a \gls{MILP} approach, to develop a global Swiss energy model. It is designed to contrast the outcomes from energy systems optimized solely for economic benefits against those optimized for reducing environmental impacts through \gls{LCA} indicators. This involves a suite of comprehensive, independent impact indicators (Areas of Concern and Areas of Protection). In 2020, the project aims to explore Switzerland's capacity to transition to an energy-independent state, focusing on broader environmental impacts, avoiding shifts in environmental burden, and highlighting potential co-benefits.

\paragraph{Research Objectives}
The core objectives of this project are:
\begin{itemize}
    \item To systematically and consistently integrate \gls{LCA} indicators within the \gls{MILP} energy system model, ensuring optimization of these environmental indicators individually.
    \item To establish a multi-objective optimization methodology that effectively assesses the trade-offs between \gls{LCA} and economic indicators in the realm of energy systems modeling.
    \item To implement \gls{LCA} modeling specifically tailored to the Swiss energy system context.
\end{itemize}

\paragraph{Research Focus}
To address these challenges, our research will focus on:
\begin{itemize}
    \item Developing techniques for accurate characterization of technologies within the \gls{LCA} framework.
    \item Formulating strategies to integrate \gls{LCIA} seamlessly into energy system modeling.
    \item Innovating methods for the monitoring and evaluation of \gls{LCA} indicators within the energy modeling process.
    \item Assessing potential shifts in environmental burdens within a transitioning low-carbon Swiss energy system.
    \item Analyzing the impact of \gls{MOO} on energy system configurations, particularly in balancing environmental and economic objectives.
\end{itemize}

Through this research, we aim to bridge the identified gaps in current literature, contributing to a more holistic and sustainable approach in energy system design and optimization.

\clearpage
\section{Methods}

\subsection{Modeling Framework}

This work builds on an exploratory work \cite{brunReducingGreenhouseGas2022,schnidrigIntegrationLifeCycle2023} integrating \gls{LCA} in the pre-existing MILP-based EnergyScope framework, a model initially formulated by Moret et al. \cite{moret_strategic_2017}, and continuously improved by Li et al. \cite{liDecarbonizationComplexEnergy2020}, and Schnidrig et al. \cite{schnidrigRoleEnergyInfrastructure2023}.

\begin{figure}[htb]
	\centering
	\includegraphics[width=0.9\textwidth]{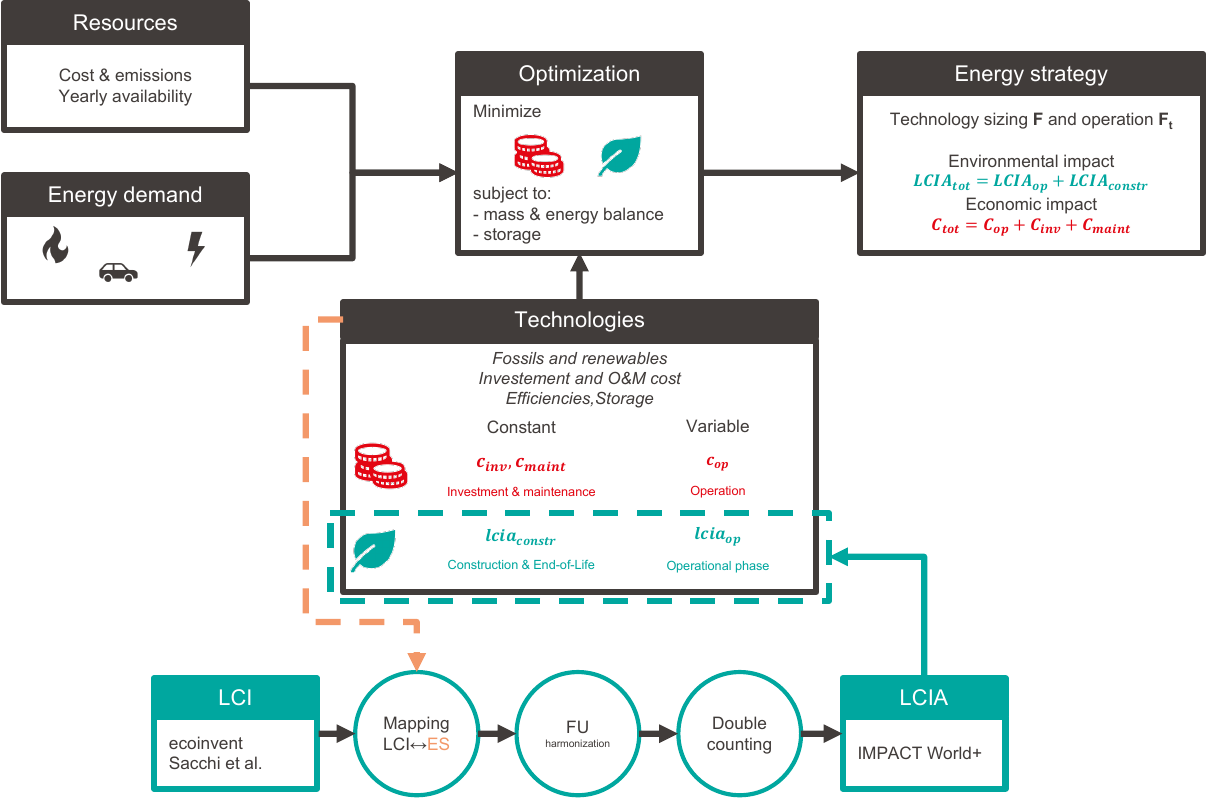}
	\caption{Graphical representation of the methodology followed integrating \gls{LCIA} indicators into \gls{ES}. The green steps at the bottom of the figure illustrate the adaptation of the \gls{LCI} to the \gls{ES} technologies to be split into variable and constant impact to allow the optimization of economic (red) and environmental (green) variables. }
	\label{fig:methodology}
\end{figure}

The \textit{EnergyScope} framework delineates a comprehensive multi-energy model, evaluated on a monthly averaged basis. This ensures a balanced mass and energy conservation equation between the demands and resources. Demand has been segmented into sectors, such as households, services, industry, and transportation, and further broken down by energy types to create a more precise categorization. In a bid to achieve optimal configurations, the key decision variables, represented by the installed size $\boldsymbol{\mathrm{F}}$ and use $\boldsymbol{\mathrm{F_{t}}}$ of technologies (Figure \ref{fig:methodology}).

\subsubsection{Economic objective}

Central to the primary \gls{OF} of EnergyScope is the total cost $\boldsymbol{\mathrm{C_{tot}}}$ (Eq. \ref{eq:ctot}). This total cost is composed of a constant and a variable part. The constant cost corresponds to the annualized technology-specific cost for investment $\boldsymbol{\mathrm{C_{inv}}}$ (Eq. \ref{eq:cinv}) and maintenance $\boldsymbol{\mathrm{C_{maint}}}$ (Eq. \ref{eq:cmaint}), defined by the decision variable $\boldsymbol{\mathrm{F}}$. As we consider only the hypothetical optimal energy system for 2020 within this study, we only consider new installations, subtracting existing infrastructure $f_{ext}$. The variable part corresponds to the use of technologies $\boldsymbol{\mathrm{F_t}}$ resulting in the operational costs of resources $\boldsymbol{\mathrm{C_{op}}}$ (Eq. \ref{eq:cop}).

\begin{alignat}{2}
    \boldsymbol{\mathrm{C_{tot}}}&=\sum_{tec}(\boldsymbol{\mathrm{C_{inv}}}(tec) \cdot \boldsymbol{\tau} (tec) + \boldsymbol{\mathrm{C_{maint}}}(tec)) + \sum_{res}\boldsymbol{\mathrm{C_{op}}}(tec) \label{eq:ctot}\\
        &\boldsymbol{\mathrm{C_{inv}}}(tec)=c_{inv}(tec) \cdot (\boldsymbol{\mathrm{F}}(tec)-f_{ext}(tec)) \label{eq:cinv}\\
        &\boldsymbol{\mathrm{C_{maint}}}(tec)=c_{maint}(tec) \cdot \boldsymbol{\mathrm{F}}(tec) \label{eq:cmaint}\\
        &\boldsymbol{\mathrm{C_{op}}}(tec)=\sum_{t} c_{op}(tec) \cdot \boldsymbol{\mathrm{F_t}}(tec,t) \cdot t_{op}(t) \label{eq:cop}\\
        &\quad \forall \:tec\in \mathcal{TEC}, \: t\in\mathcal{PERIODS},   \nonumber
\end{alignat}

\subsubsection{LCA objectives}

The environmental \gls{OF} variable $\boldsymbol{\mathrm{LCIA_{tot}}}(i)$ assesses the sum of life cycle impacts of a technology throughout its construction, operation and end-of-life (Eq. \ref{eq:lcia_tot}) for the different \gls{LCA} indicators $i$ listed in Table \ref{tab:impact_indicators}.

\paragraph{Constant impact}
The \gls{LCA} impact score $\boldsymbol{\mathrm{LCIA_{stat}}}$ compute the \gls{LCA} annualized impact score of the technology over its useful life time, encompassing its construction and its end-of-life. It is calculated by multiplying the installed technology size $\boldsymbol{\mathrm{F}}$ with the specific impact factor  $lcia_{stat}$ by unit of technology installed (Table \ref{tab:unit}), divided by the technology's useful lifetime (Eq. \ref{eq:lcia_constr}).

\paragraph{Variable impact}
The variable \gls{LCA} impact score $\boldsymbol{\mathrm{LCIA_{var}}}$ provides the impacts generated during the operation of the technology, calculated by multiplying the annual use of the technology $\sum_{t} \boldsymbol{\mathrm{F_t}}(tec,t) \cdot t_{op}(t)$ with the specific operational impact factor  $lcia_{var}$ (Eq. \ref{eq:lcia_op}). The latter is expressed in terms of impact per unit of provided energy service (impact/FU).

\begin{alignat}{2}
    \boldsymbol{\mathrm{LCIA_{tot}}}(i)&=\sum_{tec}(\boldsymbol{\mathrm{LCIA_{stat}}}(i,tec) + \boldsymbol{\mathrm{LCIA_{var}}}(i,tec)) \label{eq:lcia_tot}\\
        &\boldsymbol{\mathrm{LCIA_{stat}}}(i,tec)=lcia_{stat}(i,tec) \cdot \boldsymbol{\mathrm{F}}(tec)  \cdot \frac{1}{\boldsymbol{n} (i)} \label{eq:lcia_constr}\\
        &\boldsymbol{\mathrm{LCIA_{var}}}(i,tec)= lcia_{var}(i,tec) \cdot \sum_{t} \boldsymbol{\mathrm{F_t}}(tec,t) \cdot t_{op}(t) \label{eq:lcia_op}\\
        &\quad \forall \quad i \in \mathcal{INDICATORS}, \:tec\in \mathcal{TEC}, \: t\in\mathcal{PERIODS},   \nonumber
\end{alignat}

\begin{table}[ht]
\centering
\caption[Physical units used in ES]{Physical units of the \gls{ES} technologies' operation and construction products, depending on the end use category it belongs to.\newline
The functional units for generating LCA metrics are \textit{"Production of \SI{1}{[Operation \ unit]} in Switzerland in 2020"} for the operation phase, and \textit{"Provide \SI{1}{[Construction \ unit]} capacity in Switzerland in 2020"} for construction phase.}
\begin{tabular}{ccc}
\toprule
\textbf{End use categories} & \textbf{Operation}  & \textbf{Construction}\\ \midrule
Electricity                 & [GWh]               & [GW]\\
Heat                        & [GWh]               & [GW]\\
Mobility freight            & [Mtkm]              & [$\frac{\text{Mtkm}}{h}$]\\
Mobility passenger          & [Mpkm]              & [$\frac{\text{Mpkm}}{h}$]\\ \bottomrule
\end{tabular}
\label{tab:unit}
\end{table}

\clearpage

\subsection{Life Cycle Assessment (LCA)}

\subsubsection{Mapping of Technologies with Life-Cycle Inventories}

\begin{figure}[htb]
    \begin{subfigure}[b]{0.49\textwidth}
         \centering
	\includegraphics[width=1\textwidth]{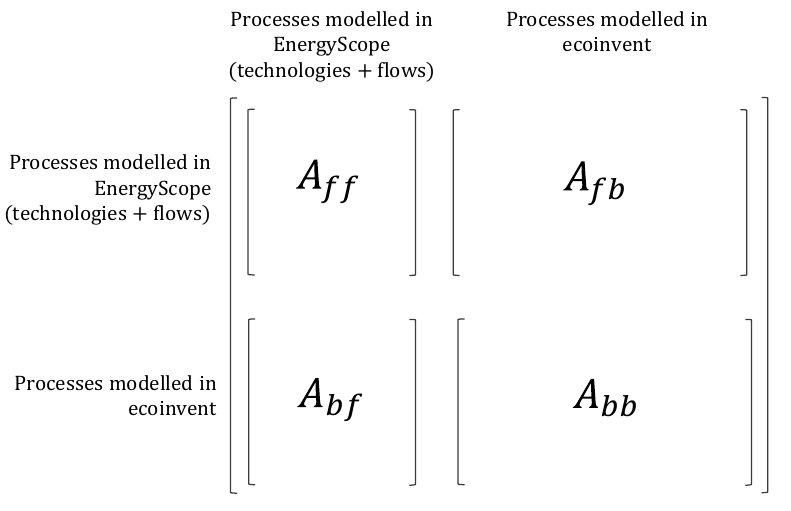}
    \caption{Matrix $A$}
    \label{sfig:matrix_A}
    \end{subfigure}
        \begin{subfigure}[b]{0.49\textwidth}
         \centering
	\includegraphics[width=1\textwidth]{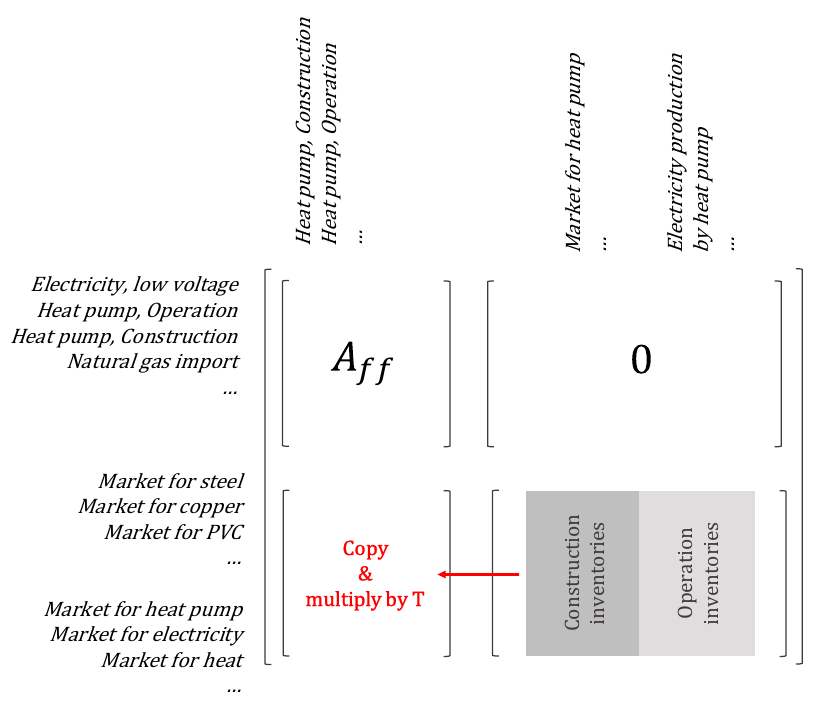}
    \caption{Mapping and data harmonization}
    \label{sfig:matrix_mapping}
    \end{subfigure}
        \begin{subfigure}[b]{0.49\textwidth}
         \centering
	\includegraphics[width=1.2\textwidth]{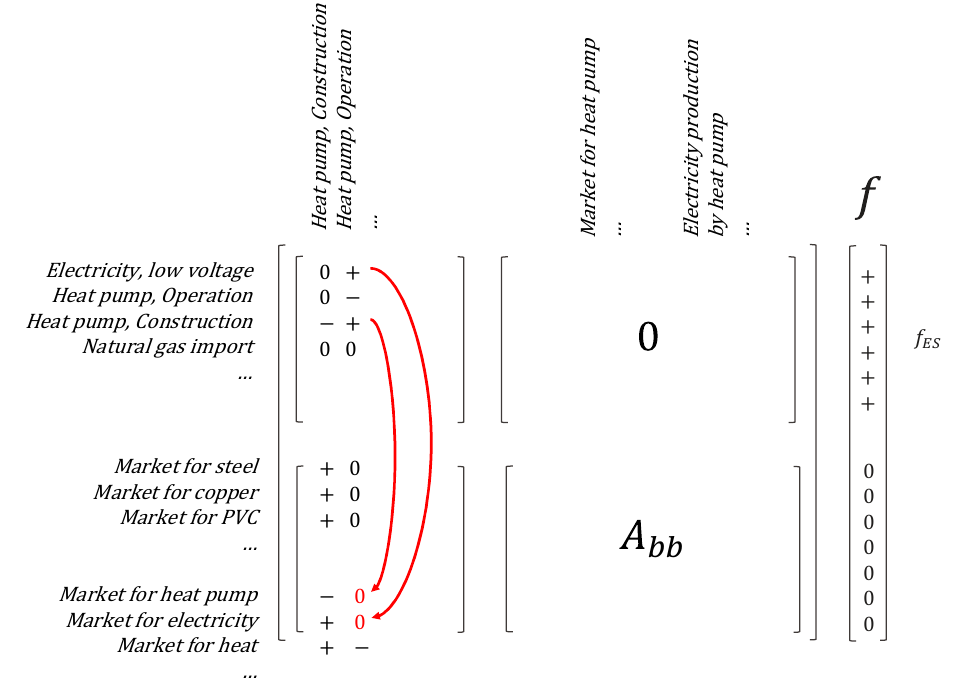}
    \caption{Double counting removing}
    \label{sfig:matrix_doublecounting}
    \end{subfigure}
    \caption[LCA technology matrix modification]{Matrix representation of the coupling between \gls{ES} and \gls{LCA}. The background-background matrix $A_{bb}$ is the original technology matrix from ecoinvent 3.8. The latter is extended as we create new processes to model \gls{ES} technologies, thus creating the overall technology matrix $A$. $A_{bb}$ is a square matrix if size $n$, where $n$ is the number of processes within ecoinvent 3.8 database. Each mapped technology generates an additional column and row. The foreground-foreground matrix $A_{ff}$ denotes the inputs (flows and construction) and outputs (flows) among \gls{ES} technologies and flows. $A_{ff}$ is a square matrix of size $n'$, where $n'$ is the number of technologies and flows in EnergyScope. Consequently, the overall technology matrix $A$ is a square matrix of size $n+n'$. $A_{bf}$ is composed of a set of columns of $A_{bb}$, identified through the mapping of technologies with ecoinvent life-cycle inventories. $A_{bf}$ columns are further multiplied by conversion factors contained in the transformation matrix $T$. In a nutshell, $A_{bf} = A_{bb} \times T$. Some entries of $A_{bf}$ are set to zero when their \gls{ES} equivalent is positive in $A_{ff}$ to avoid double-counting.}
    \label{fig:matrix_A}
\end{figure}

The systematic integration of LCA within the framework of \gls{ES} encompasses a comprehensive mapping of 216 energy conversion and storage technologies with similar technologies documented in life cycle assessment databases. Each mapped technology from the LCA databases is decomposed in a constant (comprising construction and decommissioning) and a variable life cycle stage (covering operation). For instance, in the case of \gls{PV} technology, the construction phase includes the activity "\textit{Production of \SI{1}{kW} of PV panels}", whereas the operation phase encompasses the "\textit{Production of \SI{1}{MWh} of electricity by PV panels}." Certain technologies, such as electrolysis, are devoid of an operation inventory due to their exclusive environmental impact through the consumption of \gls{ES} flows, thus eliminating redundancy (refer to Section \ref{sec:double_counting2} for details). The life-cycle inventory data is primarily sourced from ecoinvent 3.8 cut-off version (for established technologies) and from contemporary research (e.g., Sacchi et al.\cite{sacchi2022}) for emergent technologies. This data informs the expansion of the technology matrix $A$ (Figures \ref{sfig:matrix_A} and \ref{sfig:matrix_mapping}) and the elementary flows matrix $B$ in the context of \gls{ES}.

\subsubsection{Harmonization of EnergyScope and LCA Data}

The harmonization process addresses the discrepancy in units between \gls{ES} technologies operation and construction, and their life-cycle inventories activities, employing conversion factors to standardize impact measurement (detailed in Appendix \ref{tab:app_tech_map}). The background-foreground technology matrix $A_{bf}$ is the result of the multiplication between the transformation matrix $T$, which is populated by the conversion factors, and the sub-matrix of $A_{bb}$, which provides ecoinvent background processes (Figure \ref{sfig:matrix_mapping}). Conversion factors in the transformation matrix T for mobility and energy transmission technologies includes other key parameters such as occupancy rates and transmission distances.

\subsubsection{Double-counting Removal} \label{sec:double_counting2}

Double-counting removal involves identifying and nullifying redundant flows within the integrated system. This process begins by identifying all input flows of a given technology in \gls{ES} and cross-referencing them with the variable input flows from the corresponding technology in the technosphere matrix $A_{bb}$ (essentially operation and maintenance). The intersection of both lists, facilitated by the Central Product Classification (CPC), highlights double-counted flows from the operation and maintenance that are excluded from the $A_{bb}$ matrix (Figure \ref{sfig:matrix_doublecounting}). Markets denote a specific kind of ecoinvent process, as they typically average different processes leading to the same product within a region. Market are thus composed of multiple similar flows. A recursive algorithm is employed to delve into the process tree and isolate pertinent flows. This refinement is reflected in the adjusted technology matrix, denoted as $A^{*}$ (Eq. \ref{eq:h2}). 

\subsubsection{Life-cycle Impact Assessment}

The LCA metrics are derived using the \gls{IW+} v2.0.1 impact assessment method, characterized by its matrix $C$ (Eq. \ref{eq:h2}) reporting characterization factors for given impact categories. This method delineates three Areas of Concern (AoC)—carbon footprint, fossil and nuclear energy use, and water scarcity footprint—alongside the Areas of Protection (AoP), namely human health and ecosystem quality. These two latter AoP aggregate the contribution of impact categories affecting human health such as (name few of them) and respectively ecosystem quality (such as "name few of them"). The \gls{IW+} expert version allow to disaggregate these two AoP into 21 impact categories and evaluating their respective relative contribution. The computation of the overall environmental impact scores of each technology employs matrix $R$, comprising impact results vectors $h_{j}$ specific a final demand vector $f_{j}$ (Eq. \ref{eq:f_tech2}) for both operational and construction phases. In other words, an LCA is performed for each technology and phase of the model. 

\begin{equation}
    h_{j} = C \cdot B \cdot (A^{*})^{-1} \cdot f_{j}
    \label{eq:h2}
\end{equation}

\begin{equation}
    f_{j} = 
    \begin{cases}
    1, & \text{if } i = j = (tech,phase) \\
    0, & \text{otherwise}
    \end{cases}
    \label{eq:f_tech2}
\end{equation}

\clearpage
\begin{table}[!htb]
\centering
\caption[List of \gls{LCA} indicators]{\gls{LCA} indicators according to \gls{IW+} integrated in \gls{ES} with their respective abbreviation and unit.}
\begin{tabular}{c|cllc}
\toprule
\multicolumn{1}{l}{Category} & AoP & Acronym & Indicator & Unit \\\midrule
\parbox[t]{2mm}{\multirow{5}{*}{\rotatebox[origin=c]{90}{Impact profile}}} &  & CF & Carbon footprint & $\mathrm{[kg \cdot CO_2^{eq} \; (short)]}$ \\
 &  & FNEU & Fossil and nuclear energy use & $\mathrm{[MJ\; deprived]}$ \\
 &  & REQD & Remaining Ecosystem quality damage & $\mathrm{[DALY]}$ \\
 &  & RHHD & Remaining Human health damage & $\mathrm{[PDF \cdot m^2 \cdot yr]}$ \\
 &  & WSF & Water scarcity footprint & $\mathrm{[m^3\; world-eq]}$ \\\midrule
\parbox[t]{2mm}{\multirow{29}{*}{\rotatebox[origin=c]{90}{Impact categories}}} & \parbox[t]{2mm}{\multirow{12}{*}{\rotatebox[origin=c]{90}{Human Health}}} & CCHHL & Climate change, human health, long term & $\mathrm{[DALY]}$ \\
 & \multicolumn{1}{c}{} & CCHHS & Climate change, human health, short term & $\mathrm{[DALY]}$ \\
 & \multicolumn{1}{c}{} & HTXCL & Human toxicity cancer, long term & $\mathrm{[DALY]}$ \\
 & \multicolumn{1}{c}{} & HTXCS & Human toxicity cancer, short term & $\mathrm{[DALY]}$ \\
 & \multicolumn{1}{c}{} & HTXNCL & Human toxicity non-cancer, long term & $\mathrm{[DALY]}$ \\
 & \multicolumn{1}{c}{} & HTXNCS & Human toxicity non-cancer, short term & $\mathrm{[DALY]}$ \\
 & \multicolumn{1}{c}{} & IRHH & Ionizing radiation, human health & $\mathrm{[DALY]}$ \\
 & \multicolumn{1}{c}{} & OLD & Ozone layer depletion & $\mathrm{[DALY]}$ \\
 & \multicolumn{1}{c}{} & PMF & Particulate matter formation & $\mathrm{[DALY]}$ \\
 & \multicolumn{1}{c}{} & PCOX & Photochemical oxidant formation & $\mathrm{[DALY]}$ \\
 & \multicolumn{1}{c}{} & TTHH & Total human health & $\mathrm{[DALY]}$ \\
 & \multicolumn{1}{c}{} & WAVHH & Water availability, human health & $\mathrm{[DALY]}$ \\\cline{2-5}
 & \parbox[t]{2mm}{\multirow{17}{*}{\rotatebox[origin=c]{90}{Ecosystem Quality}}} & CCEQL & Climate change, ecosystem quality, long term & $\mathrm{[PDF \cdot m^2 \cdot yr]}$ \\
 &  & CCEQS & Climate change, ecosystem quality, short term & $\mathrm{[PDF \cdot m^2 \cdot yr]}$ \\
 &  & FWA & Freshwater acidification & $\mathrm{[PDF \cdot m^2 \cdot yr]}$ \\
 &  & FWEXL & Freshwater ecotoxicity, long term & $\mathrm{[PDF \cdot m^2 \cdot yr]}$ \\
 &  & FWEXS & Freshwater ecotoxicity, short term & $\mathrm{[PDF \cdot m^2 \cdot yr]}$ \\
 &  & FWEU & Freshwater eutrophication & $\mathrm{[PDF \cdot m^2 \cdot yr]}$ \\
 &  & IREQ & Ionizing radiation, ecosystem quality & $\mathrm{[PDF \cdot m^2 \cdot yr]}$ \\
 &  & LOBDV & Land occupation, biodiversity & $\mathrm{[PDF \cdot m^2 \cdot yr]}$ \\
 &  & LTBDV & Land transformation, biodiversity & $\mathrm{[PDF \cdot m^2 \cdot yr]}$ \\
 &  & MAL & Marine acidification, long term & $\mathrm{[PDF \cdot m^2 \cdot yr]}$ \\
 &  & MAS & Marine acidification, short term & $\mathrm{[PDF \cdot m^2 \cdot yr]}$ \\
 &  & MEU & Marine eutrophication & $\mathrm{[PDF \cdot m^2 \cdot yr]}$ \\
 &  & TRA & Terrestrial acidification & $\mathrm{[PDF \cdot m^2 \cdot yr]}$ \\
 &  & TPW & Thermally polluted water & $\mathrm{[PDF \cdot m^2 \cdot yr]}$ \\
 &  & TTEQ & Total ecosystem quality & $\mathrm{[PDF \cdot m^2 \cdot yr]}$ \\
 &  & WAVFWES & Water availability, freshwater ecosystem & $\mathrm{[PDF \cdot m^2 \cdot yr]}$ \\
 &  & WAVTES & Water availability, terrestrial ecosystem & $\mathrm{[PDF \cdot m^2 \cdot yr]}$\\\bottomrule
\end{tabular}
\label{tab:impact_indicators}
\end{table}

\subsection{Multi-Objective Optimization (MOO)}
\gls{MOO} is an integral component of multi-criteria decision-making. It focuses on solving mathematical problems where several \glspl{OF} must be optimized simultaneously. This approach is especially beneficial in energy planning scenarios with multiple competing objectives. For instance, EnergyScope, in its traditional applications, either optimized a single \gls{OF} \cite{moret_strategic_2017,limpensEnergyScopeTDNovel2019,liDecarbonizationComplexEnergy2020} or engaged in bi-objective optimization, generating carbo-economic Pareto curves in specific cases \cite{schnidrigModellingFrameworkAssessing2021}.

\gls{MOO} aims to achieve various goals, which include:
\begin{itemize}
    \item Identifying solutions that align with the specific preferences of a decision-maker.
    \item Curating a diverse set of Pareto-optimal solutions.
    \item Quantifying the necessary trade-offs associated with each objective.
\end{itemize}

In this research, we have approached \gls{MOO} by incorporating the \gls{LCIA} OF, denoted as $l\in \mathcal{LCIA-I}\subset\mathcal{OF}$, as constraints during economic optimization (as shown in Eq. \ref{eq:f_obj_moo}). This strategy defines the size and use of technology denoted as $\boldsymbol{\mathrm{F}}$ and $\boldsymbol{\mathrm{F_t}}$, respectively. The weighting mechanism, provided in Eq. \ref{eq:fobj_constr}, navigates through the multidimensional Pareto curve, distinguishing between the extreme points recognized during \gls{SOO}. Here, $\omega(j)=1$ and $\omega(i)=0, \; \forall i \in \mathcal{OF}\setminus \{j\}$.

\begin{alignat}{2}
    \min_{\boldsymbol{\mathrm{F}},\boldsymbol{\mathrm{F_{t}}}} \; & \boldsymbol{\mathrm{C_{tot}}}
    \label{eq:f_obj_moo}\\
                & \mathrm{s.t.}\quad \boldsymbol{\mathrm{f_{obj}}}(i)\leq \omega(i) \cdot f_{obj}^{\max}(l) + (1-\omega(i)) \cdot f_{obj}^{\min}(l)\label{eq:fobj_constr}\\
                &\quad \forall \quad i \in \mathcal{OF} = \mathcal{COST} \cup \mathcal{LCIA-I}\nonumber
\end{alignat}

To comprehensively explore the solution space of \gls{MOO}, we employed the quasi Monte-Carlo methodology \cite{metropolisMonteCarloMethod1949} Sobol to sample the weights $\omega$. The decision variables elucidate the modeled solution space $\boldsymbol{\mathrm{F}},\boldsymbol{\mathrm{F_{t}}}$, considering the varying weighting parameters $\omega(i)$, as presented in Eq. \ref{eq:MC}. Moreover, the probability of each weight $\omega(i)$ is uniformly distributed between 0 and 1 around their median value $\Tilde{\omega}$, as indicated in Eq. \ref{eq:MC_omega}.

\begin{alignat}{2}
    \boldsymbol{\mathrm{F}}(i),\boldsymbol{\mathrm{F_{t}}}(i) : \quad 
        & f((\boldsymbol{\mathrm{F}}(i),\boldsymbol{\mathrm{F_{t}}}(i)), \omega(i))\label{eq:MC}\\
        & \mathrm{s.t.} \quad  \omega(i) = P(\Tilde{\omega},\mathrm{U}(0,1))\label{eq:MC_omega}
\end{alignat}

\subsection{Application}
The methodology was applied to the 2020 Swiss energy system to ensure consistency in term of temporal and technological representativeness between the technologies of ES with the \gls{LCA} inventory database.  Energy demands were derived from the 2020 Swiss energy statistics \cite{leuchthaler-felberSchweizerischeGesamtenergiestatistik20192020}, with the overarching aim to explore hypotetical scenarios that meets the long-term ambition of a carbon-free energy system by 2020, as advocated by \gls{EP50+} (\cite{kemmlerEnergieperspektiven2050Technischer2021}), while ensuring energy supply security, but acknowledging the limitation of relying on the current technological context. This scenario deliberately excludes nuclear power, focusing on Switzerland's 'best case' in 2020, characterized by reliance on domestic energy sources and without energy imports to bolster supply security.

\textbf{Demands}

The energy statistics provide a breakdown of the final energy demand segmented by sectors. By leveraging the 2020 energy demand data \cite{leuchthaler-felberSchweizerischeGesamtenergiestatistik20192020} and categorizing it into distinct energy types. These sectoral demands are summarized in Table \ref{tab:eud_2050}.

\begin{table}[htb]
\centering
\caption{Estimated annual final energy demand by sector and energy type 2020.}
\begin{tabular}{lc|cccc}
\toprule
 &  & \textbf{Households} & \textbf{Services} & \textbf{Industry} & \textbf{Mobility} \\
\midrule
\textbf{Electricity LV} & $\mathrm{[GWh]}$ & 9818 & 9154 & 0 & 0 \\
\textbf{Electricity MV }& $\mathrm{[GWh]}$ & 0 & 1407 & 3173 & 0 \\
\textbf{Electricity HV} & $\mathrm{[GWh]}$ & 0 & 0 & 5350 & 0 \\
\textbf{Heat HT} & $\mathrm{[GWh]}$ & 0 & 183 & 5855 & 0 \\
\textbf{Heat LT SH} & $\mathrm{[GWh]}$ & 31849 & 6994 & 1965 & 0 \\
\textbf{Heat LT HW} & $\mathrm{[GWh]}$ & 6322 & 1605 & 393 & 0 \\
\textbf{Freight} & $\mathrm{[Mtkm]}$ & 0 & 0 & 0 & 21106 \\
\textbf{Passenger} & $\mathrm{[Mpkm]}$ & 0 & 0 & 0 & 74590\\
\bottomrule
\end{tabular}
\label{tab:eud_2050}
\end{table}

\textbf{Potentials}

In modeling our energy system, we prioritize the security of supply by reducing any energy vector imports to zero. Consequently, all primary energy sources are derived from within the region under study, as defined by their potentials (see Table \ref{tab:potential}). We model the potential of the resources from a technical perspective rather than an economic one, as this latter is subject to arbitrary and uncertain estimation of future renewable energy markets.

\begin{table}[htb]
\centering
\caption{Annual potential for resources and renewable energy technologies.\newline
Values in parentheses for hydropower technologies represent potential with reinforcement.}
\begin{tabular}{l|ccccc}
\toprule
\textbf{Resources} & \textbf{Waste Fossil} & \textbf{Waste Bio} & \textbf{Wood} & \textbf{Wet Biomass} & \textbf{Hydro Storage} \\\midrule
$\mathrm{[GWh]}$ & 10833 & 8917 & 15278 & 12472 & 8900 \\\midrule
\textbf{Technologies} & \textbf{Geothermal} & \textbf{Hydro Dam} & \textbf{Hydro River} & \textbf{PV} & \textbf{Wind} \\\midrule
$\mathrm{[GW]}$ & 4.8 & 8.08 (8.52) & 3.8 (4.65) & 67 & 20 \\
$\mathrm{[GWh]}$ & 42.08 & 17.48 & 19.726 & 66.4 & 40.3\\
\bottomrule
\end{tabular}
\label{tab:potential}
\end{table}

\section{Results}

\subsection{Single-Objective Optimization}
\begin{figure}[! htb]
	\centering
	\includegraphics[width=\textwidth]{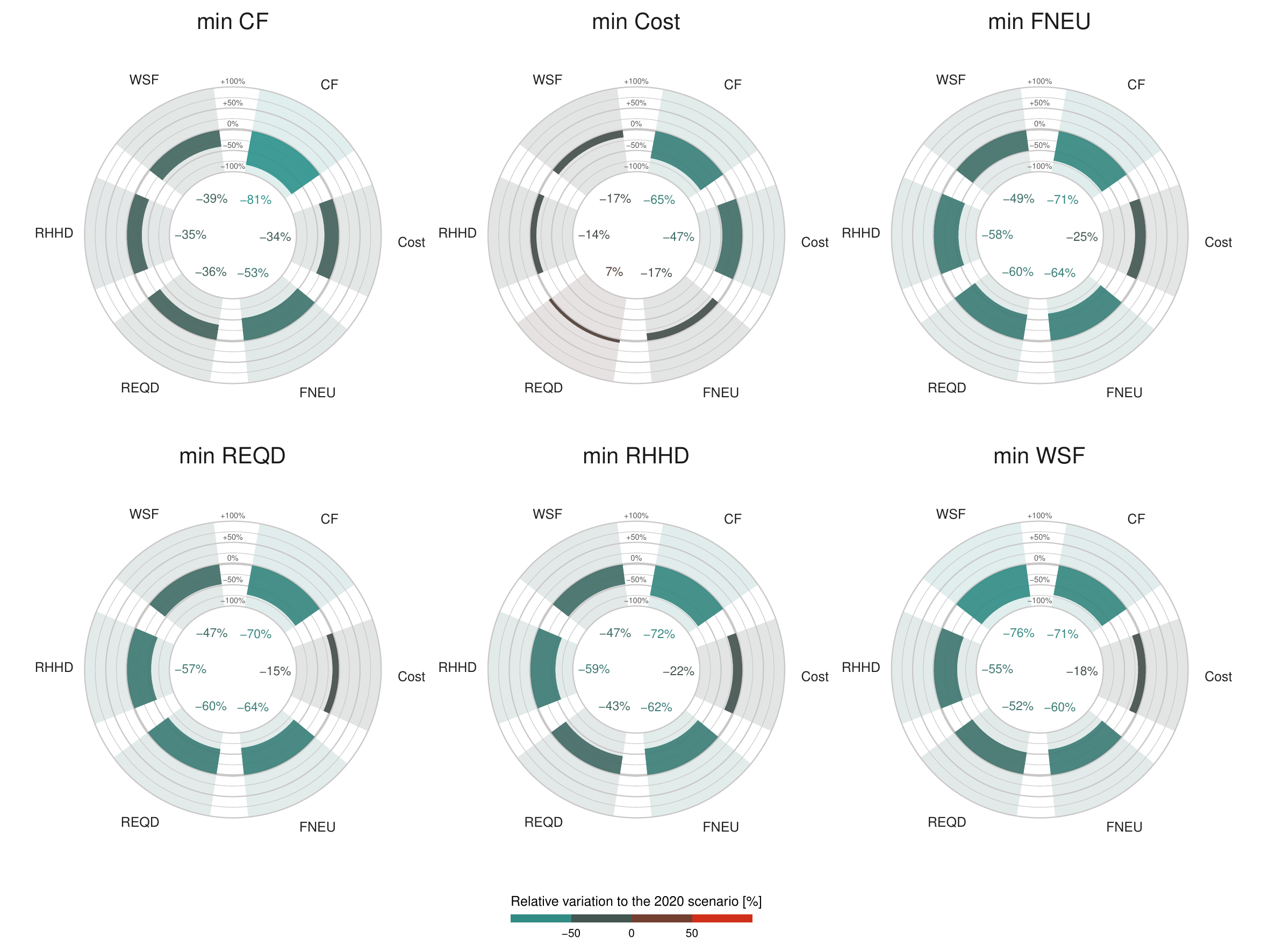}
	\caption{OF values comparison for \gls{SOO}. Each sub-figure corresponds to an individual optimization. The height of the segments corresponds to the \gls{OF}'s relative variation to the 2020 reference scenarios \glspl{OF} values [\%].\newline \textbf{\acrshort{CF}}: \acrlong{CF}, \textbf{Cost}: Total Cost, \textbf{\acrshort{FNEU}}: \acrlong{FNEU}, \textbf{\acrshort{REQD}}: \acrlong{REQD}, \textbf{\acrshort{RHHD}}: \acrlong{RHHD}, \textbf{\acrshort{WSF}}: \acrlong{WSF}}
	\label{fig:OF_spider}
\end{figure}

Pursuing of a comprehensive environomic analysis, each OF was individually optimized. This initial step was pivotal to ascertain the boundary values $f_{OF}^{\min}$ and $f_{OF}^{\max}$, which serve as foundational metrics the subsequent \gls{MOO}. A detailed representation of the tracking of \gls{OF} values through individual optimization is illustrated in Figure \ref{fig:OF_spider}. These values are evaluated to the current 2020 scenario to provide a coherent context, offering a comparative perspective on the shifts and nuances observed across each isolated optimization.

\paragraph{Overview of Optimizations and Trade-offs}

Minimizing different objective functions unveiled  trade-offs between the economic and environmental objectives. For instance, when focusing solely on the reduction of \gls{CF}, a decrease of 81\% of the same indicator was observed, accompanied by a significant drop in cost by 34\%  and reductions of the remaining environmental indicators of 35\%-53\% compared to the 2020 baseline scenario. However, when directing efforts to minimize cost, a substantial decrease in \gls{CF} by 65\% can be observed, but at the expense of a minor increase in \gls{REQD} by 7\%. This pattern underscores that while striving to minimize costs might yield environmental benefits, such as a reduced carbon footprint, but it could also unintentionally aggravate other environmental challenges.
%Inter-environmental trade-offs were discernible, too. A marked decrease in one environmental factor often correlated with significant benefits in others. For instance, minimizing \gls{CF} not only reduced cost but led to decreases in the other environmental indicators between 35\% and 53\%, suggesting that a narrow focus on carbon footprint leads to parallel benefits on other environmental dimensions. Parallel benefits in costs and environmental impacts appear in all \gls{SOO} optimizations, except for the previously described increase in \gls{REQD} in cost minimization. 

\paragraph{Diversifying the Environmental Focus}

The results emphasize the intricacy of environomic relations. While \gls{CF}, with its variations ranging between 65\% and 81\%, stands out as a significant environmental factor, our findings prompt us to consider the broader picture. This realization accentuates the necessity to expand our environmental scope, recognizing that metrics such as \gls{FNEU}, \gls{REQD}, \gls{RHHD}, and \gls{WSF}, which have shown variations as pronounced as a 76\% decrease in the case of \gls{WSF} minimization, demand our attention and can't be overlooked.

\begin{figure}[htb]
	\centering
	\includegraphics[width=0.9\textwidth]{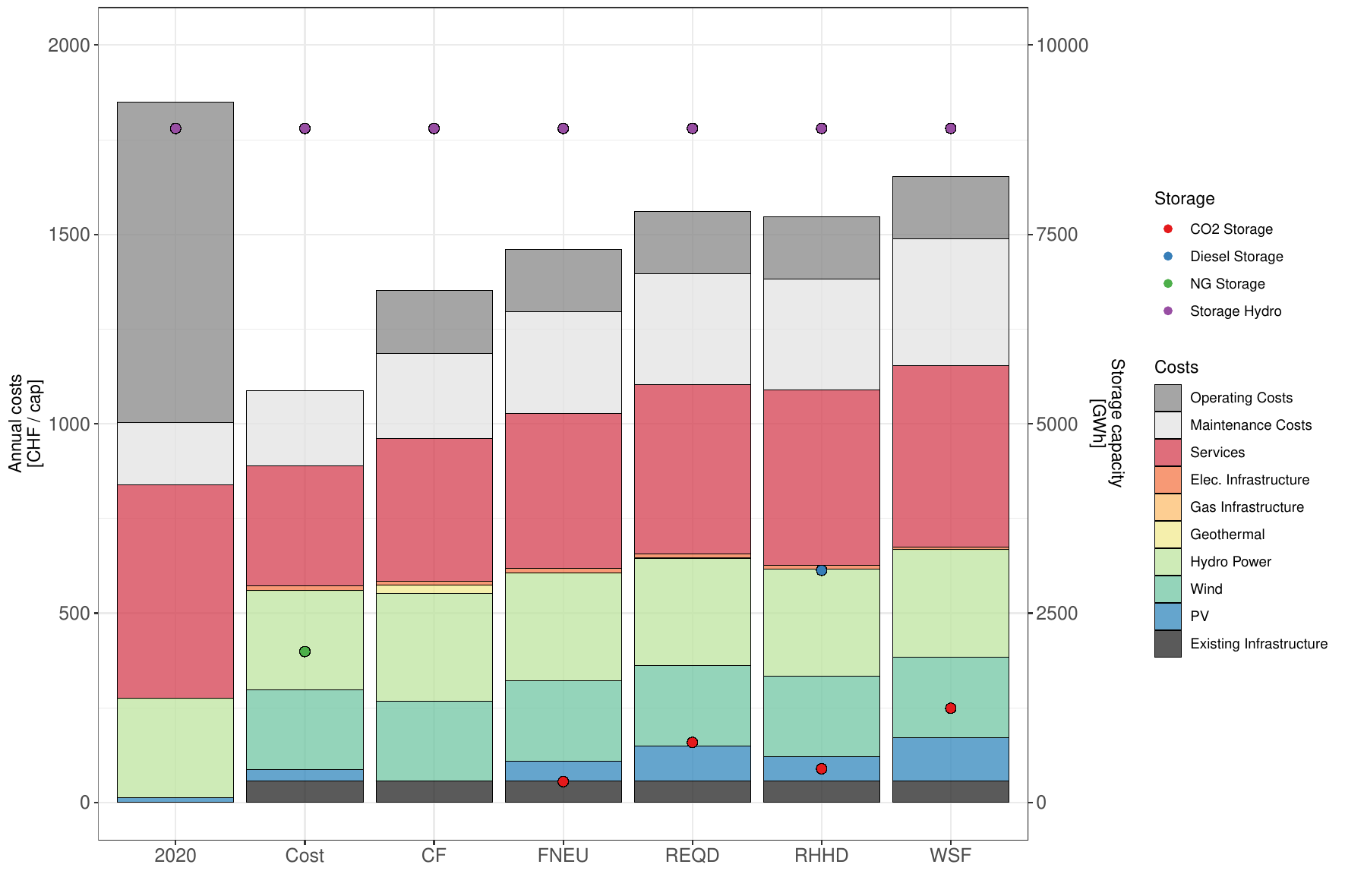}
	\caption{Overall cost composition of energy systems for single-objective optimizations. The secondary axis highlights installed storage capacity. The 2020 scenario represents the current Swiss energy system and the other six represent hypothetical scenario for an energy-independent Switzerland 2020 with single objective optimization.}
	\label{fig:capex_overview}
\end{figure}

\paragraph{Energy system configurations}
All optimal configurations, when compared to the 2020 reference scenario, have demonstrated an cost reduction of the energy system. The overall cost composition for the 2020 baseline scenario and six \gls{OF} minimizations is depicted in Figure \ref{fig:capex_overview}.
% suggestion: All optimal configurations, when compared to the 2020 reference scenario, have demonstrated a cost reduction as depicted in Figure \ref{fig:capex_overview}.

The optimal configurations are based on an energetically self-sustained system, thereby eliminating energy imports. Such optimized configurations correspond to hypothetical scenarios which are lean on local energy resources, with distinctions noticeable in utilizing primary energy sources, mobility and heating service technologies, and energy storage requisites. In contrast, the 2020 reference case representing current Swiss energy system, heavily depends on operational costs due to extensive fossil fuel imports. 

In all optimized scenarios, renewable resources are leveraged extensively as seen in Figure \ref{fig:capex_overview}. Wind turbines are deployed to their full potential, synergizing with existing hydropower plants. Variations arise when complementing this hydro-wind ensemble: \glspl{PV} span between \SI{7}{GW}-\SI{20}{GW}. An exception is the \gls{CF} minimization, which opts for geothermal power plants (\SI{3.5}{GW}) over \gls{PV}. The tangible operational costs stem from harnessing local biomass—utilized maximally in most scenarios, with the cost minimization scenario as an exception. Sectoral analysis can be found in the supplementary material \ref{app:SOO}.

%\newpage
\subsection{Multi-Objective Optimization}
\begin{figure}[!htb]
	\centering
	\includegraphics[width=\textwidth]{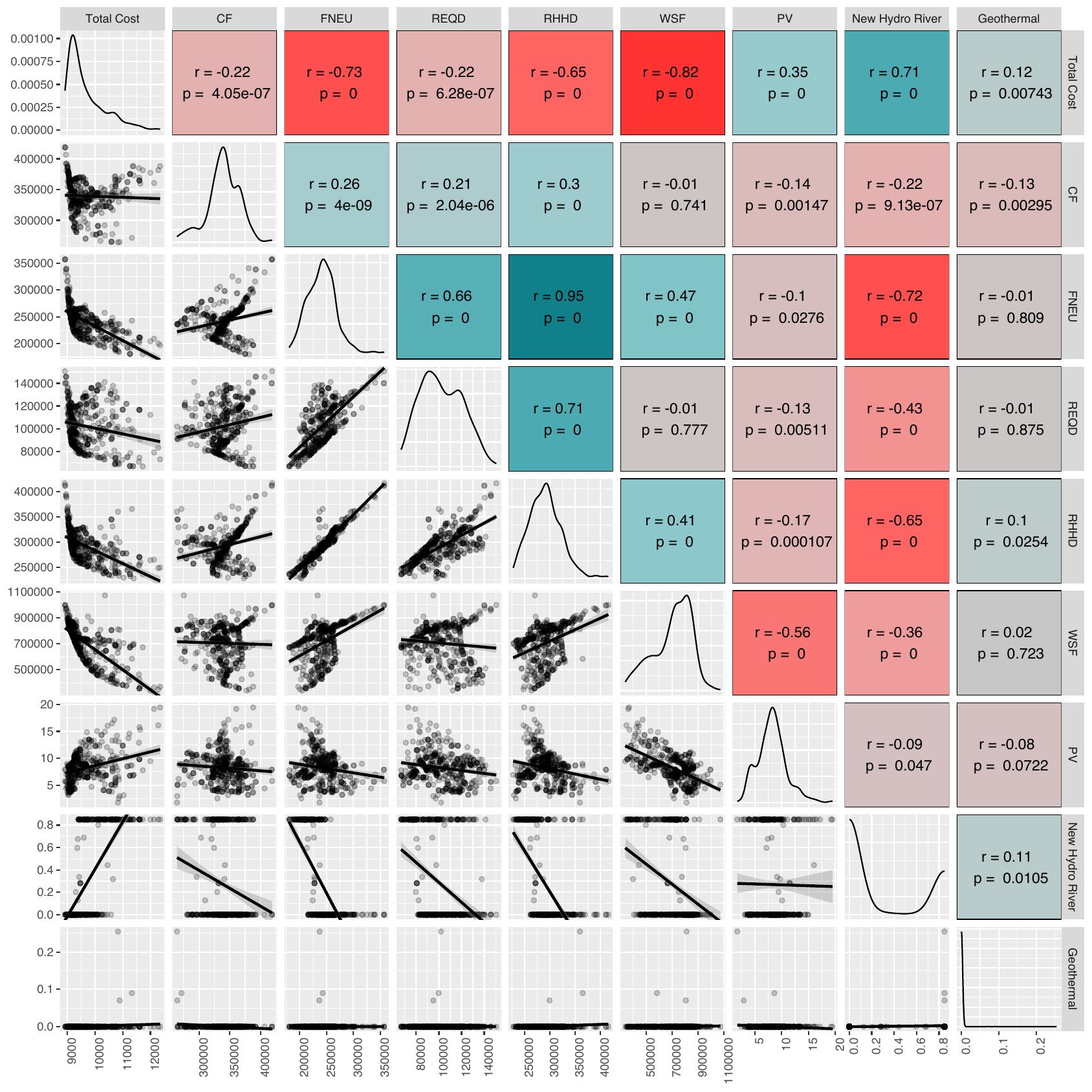}
	\caption{Pearson correlation coefficient matrix based on the \gls{MOO}. The upper triangle depicts the correlation factor $r$ with the color gradient and the significance $p$ with the transparency. The diagonal depicts the distribution of the appearance of the individual variables. The lower triangle represents the observation distribution with the corresponding trend line and confidence interval. Each point corresponds to one distinct configuration. The Pareto-front between the \gls{OF} can be observed.}
	\label{fig:pearson_RE}
\end{figure}

Upon executing the \gls{MOO}, as detailed in Equation \ref{eq:MC}, for \SI{500} iterations varying the \gls{OF}-weight $\omega$, 500 distinct Pareto-optimal configurations emerge. A comparative analysis among these configurations was performed, constructing the Pearson correlation coefficient matrix, focusing on the technologies and the \glspl{OF} (Fig. \ref{fig:pearson_RE}).

\paragraph{Technological and environomic Trade-Offs}

Distinct correlations between technologies and \glspl{OF} are identifiable within the matrix's upper segment. Notably, a strong positive correlation is present between \gls{RHHD} and \gls{FNEU}, evidenced by a correlation coefficient ($r=0.95$). This suggests a directly proportional relationship where an increase in \gls{RHHD} results in a concurrent rise in \gls{FNEU} and vice versa. Conversely, a significant negative correlation is observed between \gls{WSF} and Total Cost ($r=-0.82$), implying that reduced water usage incurs higher energy system costs, thereby restricting the employment of existing hydroelectric infrastructure and necessitating supplemental energy solutions. Lesser yet noticeable correlations are apparent between Total Cost and other environmental \glspl{OF}, specifically \gls{FNEU} ($r=-0.73$) and \gls{RHHD} ($r=-0.65$), indicating that elevated environmental impacts accompany cost reductions. Although cost is inversely related to environmental \glspl{OF}, \gls{FNEU} demonstrates positive correlations with other environmental impacts, exhibiting a strong relationship with \gls{RHHD} ($r=0.95$) and \gls{REQD} ($r=0.66$), as well as weaker associations with \gls{WSF} and \gls{CF}, thereby signaling a substantial potential for synergistic environmental benefits. \gls{RHHD} exhibits a similar pattern, with strong correlations with both \gls{FNEU} and \gls{REQD} ($r=0.71$), and weaker with \gls{CF} and \gls{WSF}.

In the context of technology and \glspl{OF} interrelations, negative correlations suggest a minimizing effect on the \gls{OF} with increased technology deployment. For example, New Hydro River demonstrates environmental benefits with negative correlations to \gls{FNEU} ($r=-0.72$) and \gls{RHHD} ($r=-0.65$), yet this is coupled with an escalated Total Cost ($r=0.71$). \gls{PV} presents a parallel trend; elevating Total Cost with increased deployment ($r=0.35$), while concurrently diminishing \gls{WSF} ($r=-0.56$). Geothermal technology exhibits only nominal correlations with the remaining \glspl{OF} and technologies, indicating a less pronounced impact on the overall system.

The scatter plot divulges the assortment of solutions in the matrix's lower part. The Pareto frontier concerning environmental indicators is discernible within the Total Cost vector. Deviations from this frontier primarily emerge from solutions with a diminished focus on cost optimization. The robust correlation between \gls{RHHD} and \gls{FNEU} is prominent, with data points closely clustered along the linear regression line. A thorough examination of patterns in the \gls{LCA} indicator columns reveals a wide dispersion of data points attributed to the stochastic nature of the Monte Carlo simulations (Eq. \ref{eq:MC_omega}).

\paragraph{Exploring Renewable Energy Synergies: Correlations and Trade-offs in \gls{MOO}}

In the elucidation of Figure \ref{fig:pearson_RE}, a detailed exposition is conducted with the coordinates X and Y, symmetrically representing the objective functions, predominantly the costs and areas of concern juxtaposed against key renewable technologies like PV, new hydro river, and geothermal energy. Notably absent is wind energy, given its consistent deployment at maximal capacity. Each point delineated within the matrix symbolizes an optimization instance with varied weighting among the objectives, all of which are situated on a 6-dimensional Pareto front. This matrix distinctly excludes the current energy system, recognizing its inefficiency in meeting any of the considered objectives. A specific case in point is the analysis of the indicators \gls{FNEU} and \gls{RHHD}. As discerned in the matrix segment at (\gls{RHHD}/\gls{FNEU}), these objective functions demonstrate a significant correlation ($r=0.95$, denoted by a green square), which is further elucidated through the point distribution and trend line in the segment (FNEU/\gls{RHHD}). This correlation suggests that increases in \gls{RHHD} are paralleled by increases in \gls{FNEU} and vice versa. For instance, minimizing \gls{RHHD} concurrently minimizes \gls{FNEU}. This trend is further illuminated by examining the correlation of \gls{RHHD} with specific technologies. A prime example is the inverse relationship between New Hydro River and \gls{FNEU} ($r=-0.72$, indicated by a red square), suggesting that an increase in New Hydro River installations correlates with a decrease in \gls{FNEU}. However, it is imperative to approach these findings with caution. As all points lie within the optimal space, the trends are not universally applicable but confined to the solution space. Thus, the assertion that increased New Hydro River installations invariably lead to lower \gls{FNEU} values is not a generalizable conclusion but a trend observed within this specific set of optimal solutions.

\paragraph{Technology size distribution}
The single-peak distribution, indicative of a single optimal systemic value, is seen for Total Costs (with the highest frequency at \SI{10.5}{\%} occurring at \SI{9500}{MCHF/year}), \gls{FNEU} (\SI{8.4}{\%}), and \gls{RHHD} (\SI{7.4}{\%}). These distributions suggest a preferred optimal solution across optimizations. The distance of these peaks from the respective minima in environomic indicators underscores the inherent trade-offs among the objective functions, evidenced by a relatively low frequency (\SI{4}{\%}) for the minimal Total Cost configuration, highlighting that none of the indicators is positioned at its potential minimum.

Multi-peak distributions are observed for \gls{REQD}, which shows a bimodal distribution, and for \gls{CF} and \gls{WSF}, each presenting a primary peak accompanied by a less pronounced secondary peak. Notably, \gls{PV} installation capacity exhibits a primary peak at \SI{8}{GW}, with subsequent lesser peaks at \SI{5}{GW} and \SI{12}{GW}.

Extreme behavior, characterized by a binary distribution pattern of either zero installation or maximum potential utilization, is noted in the installed capacities of renewable technologies. New Hydro River installations, for instance, are either absent (\SI{79.5}{\%} of cases) or realized at the full potential of \SI{0.8}{GW} in the remaining instances. An even more pronounced extreme distribution is observed for geothermal power plants, which are installed at the available potential of \SI{0.25}{GW} in only \SI{1.2}{\%} of cases. Unrepresented renewable resources such as Wind (\SI{20}{GW}), Hydro Dams (\SI{8.08}{GW}), and Hydro River (\SI{3.8}{GW}) also exhibit extreme behavior, as they are consistently utilized at their maximum capacity across all scenarios, thereby denoting a characteristic of objective-independent optimality in the optimization process.

%%\newpage

\section{Discussion}
\paragraph{Burden-Shifting}
When minimizing total costs of the 2020 the Swiss energy system, \gls{CF} is notably reduced by \SI{65}{\%}, reflecting a significant synergy between cost and carbon footprint reduction. However, this focus on cost tends to exacerbate other environmental indicators, manifesting a burden shift rather than a net reduction in environmental impact, as shown in similar economic optimizations of the carbon-neutral Switzerland \cite{schnidrigRoleEnergyInfrastructure2023}.
Furthermore, the data reveals that specific environmental dimensions, especially \gls{FNEU} and \gls{REQD}, are highly reactive to fluctuations in cost. This insight is crucial as it suggests that economic interventions might occasionally lead to environmental ramifications not immediately apparent if success is judged by the decrease of the carbon footprint alone. Conclusively, a comprehensive and diversified approach, rather than a carbon-centric paradigm, is indispensable for genuine environomic synergy.

Conversely, targeting \gls{CF} minimization yields broad environmental improvements, with only slight increases observed in \gls{RHHD} and \gls{REQD}. This suggest a more balanced trade-off between different environmental aspects.

Crucially, targeting the minimization of other environmental indicators invariably results in better overall environomic outcomes. These strategies surpass the current system in both environmental and economic terms, underscoring the benefit of a multi-faceted environmental approach to energy system optimization. The findings advocate for nuanced strategies that consider the interconnectedness of environomic indicators to avoid shifting burdens and achieve genuine sustainability improvements.

\paragraph{Renewable is the solution, but...}
Switzerland's energy system leverages wind and hydro resources to their limits, with wind installations at \SI{20}{GW} and hydro at \SI{8.08}{GW} across all examined scenarios. Solar PV's capacity fluctuates between \SI{7}{GW} and \SI{20}{GW}, indicating adaptive strategies for solar energy use based on varying environmental and economic goals.

Energy storage is a linchpin in managing the supply-demand balance, with hydro dam storage consistently maxed at \SI{8.9}{TWh}. Scenarios also integrate \SI{2}{TWh} of biogenic methane or \SI{3.5}{TWh} of diesel for seasonal energy storage, underscoring the tailored approaches to achieving system resilience and environmental goals.

\paragraph{Dependencies}
%MOO indicates correlations between indicators and technologies
The study shows that environmental and economic goals in energy systems are closely interconnected, often with direct trade-offs between cost and environmental impacts. Specifically, the positive correlation between \gls{RHHD} and \gls{FNEU} with $r=0.95$ indicates that minimizing the reliance on fossil fuels and nuclear energy can significantly reduce damages on human health \gls{RHHD}   and vice-versa. Conversely, the negative correlation between \gls{WSF} and Total Cost ($r=-0.82$) reveals the economic challenges inherent in reducing water use, as less reliance on hydroelectric power demands investment in alternative, more expensive energy sources such as geothermal power plants.

The study also highlights the complexities in the energy transition, where choosing environmentally beneficial technologies may result in higher costs. For example, adopting new hydro river technology demonstrates environmental benefits but at the expense of increased total costs ($r=0.71$ with Total Cost), illustrating the difficulty of simultaneously achieving both economic and comprehensive environmental goals. The nuanced relationship between Total Cost and environmental objectives is evident, where cost reductions may potentially increase environmental impacts ($r=-0.73$ with \gls{FNEU} and $r=-0.65$ with \gls{RHHD}), suggesting that low-cost energy solutions could have unintended environmental consequences. Nevertheless, the study demonstrates that minimizing each environmental indicator leads to a more cost-effective solution than in 2020.

\paragraph{Towards a renewable energy system}
% distributions
In analyzing Swiss energy system optimization, single-peak distributions for Total Costs, \gls{FNEU}, and \gls{RHHD} converge towards specific system values, yet they remain suboptimal. For instance, the highest frequency for Total Costs occurs at 9,500 MCHF (10.5\%), suggesting a generally accepted system cost that is not at the absolute minimum, highlighting necessary trade-offs between objectives. Multi-peak distributions in indicators like \gls{REQD}, \gls{CF}, and \gls{WSF}, with \gls{PV} installations peaking at \SI{8}{GW} and showing lesser peaks at \SI{5}{GW} and \SI{12}{GW}, imply multiple potential configurations that cater to different prioritizations, enforcing the inevitability of compromise in system design.

Extreme binary distributions in technology installations further illustrate this point. For example, New Hydro River installations are either fully utilized at \SI{0.8}{GW} or not at all, evident in \SI{79}{GW} of cases. Similarly, geothermal plants reach their \SI{0.25}{GW} potential in only \SI{1.2}{GW} of cases. On the other hand, technologies like Wind (20 GW), Hydro Dams (\SI{8.08}{GW}), and Hydro River (\SI{3.8}{GW}) are consistently maximized, indicating a strategic preference irrespective of the scenario. This showcases the system's inclination towards certain technologies, underscoring a deliberate trade-off in exploiting renewable resources to meet set environomic objectives.

\paragraph{Go green: A Perspective on Burden-Shifting}
The phenomenon of burden-shifting in the context of energy system optimization presents a nuanced landscape, particularly when contrasting \gls{SOO} with \gls{MOO} results. In the realm of \gls{SOO}, we observe clear instances of burden-shifting: for instance, minimizing total cost significantly enhances the \gls{CF} but at the expense of other environmental indicators. This points towards a unilateral approach where focusing excessively on one objective can inadvertently aggravate other environmental facets. However, the landscape shifts when considering MOO. Here, positive correlations emerge across all environmental indicators, indicating a more harmonious and balanced improvement across various environmental aspects. This dichotomy between \gls{SOO} and \gls{MOO} underlines the pitfalls of extreme, singular-focus strategies in optimizing energy systems.

This observation aligns with the current state of energy systems, which are far from their optimal states. In such a scenario, improving any single environmental indicator does not lead to a negative impact on others. This underscores the importance of a balanced, multi-dimensional approach to optimization, where improvements are not just desirable but necessary in any direction. By doing so, it's possible to navigate away from the pitfalls of burden-shifting and move towards a more sustainable and holistic optimization of energy systems, aligning with the broader goals of environmental stewardship and sustainable development.

\paragraph{Limitation}
The study's discussion reveals several key limitations that influence the interpretation of our results. 

Firstly, focusing on Switzerland in 2020, the case study's reliance on the EcoInvent database, which assesses the current energy system, might limit the generalization of our findings to other geographical regions or different time periods. 

Another significant limitation is the absence of regionalization in impact value assessments. This methodological choice led to the application of global average values for certain parameters, most notably in the water scarcity footprint. By using a global impact factor, hydro power stations in Switzerland, where water scarcity is a minor issue compared to the global average, were inadvertently disadvantaged.

Additionally, the exclusion of carbon as a resource within the energy system is a notable limitation. This omission prevented the implementation of negative emission technologies, thereby hindering the creation of fully carbon-neutral scenarios that compensate for both direct and indirect emissions of the energy system.

Finally, the study's approach presents a challenge in adequately addressing the remaining areas of protection, specifically \gls{RHHD} and \gls{REQD}. By individually considering carbon footprint and water scarcity footprint, the study overlooks the significant interplay between these indicators and their collective impact on \gls{RHHD} and \gls{REQD}. This oversight suggests a need for more integrated assessment methods in future research to fully capture the interconnected impacts on these crucial areas of protection.
\section{Conclusions}
% Methodological Innovations and Technological Validation
\paragraph{Methodological Innovations and Technological Validation}
This study marks a significant advancement in integrating \gls{LCIA} impacts within energy system modeling, particularly tailored for the Swiss energy landscape. We introduced a methodological framework that not only accurately characterizes technologies within \gls{LCI} but also embeds \gls{LCA} into the energy system modeling process. The inclusion of diverse environmental metrics alongside traditional economic considerations paves the way for a more comprehensive and multifaceted approach to energy system analysis.

\paragraph{Burden shifting in Single Objective Optimization}
While any optimizations (economic and environmental) lead to cost and environmental impact reductions compared to the current energy system, environmental burden shifting in between the environmental optimal solutions appears. This outcome highlights the complexity of balancing environmental and economic objectives and the importance of considering a broader range of environmental impacts in energy system optimization.

\paragraph{Compromises in Multi Objective Optimizations}
Our study delves deeply into MOO, presenting solutions to avoid environmental burden shifting. By optimizing for multiple environmental indicators simultaneously, MOO approaches succeeded in minimizing overall environmental impacts without the trade-offs observed in single-objective optimization. For example, strategies focusing on reducing the \gls{CF} improved the overall environmental profile while causing only marginal burden shifts in \gls{RHHD} and \gls{REQD}.

\paragraph{Impact on Swiss Energy System Configuration}
The application of our methodology to the Swiss energy system optimization yields critical quantitative findings. Renewable energy sources, such as wind power, were maximized to \SI{20}{GW}, and hydro power to \SI{8.08}{GW}. Solar PV installations varied between \SI{7}{GW} and \SI{20}{GW}, depending on the optimization scenario. For energy storage, hydro dam storage was key at \SI{8.9}{TWh}, supplemented by biogenic methane (\SI{2}{TWh}) and diesel reserves (\SI{3.5}{TWh}), highlighting their role in balancing seasonal fluctuations.

\section{Future work}
Beyond identifying the optimal peaks in indicators and technologies that showcase trade-offs, future research should delineate typical configurations that energy planners can reliably refer to, which implies establishing clearer direct correlations between specific technologies and environmental indicators, aiding in designing sustainable energy systems.

While the \gls{MOO} assessment reveals correlations between environomic objective functions, there is a need to determine which environomic indicators should be prioritized for detailed monitoring. Future studies should aim to identify these critical indicators that significantly influence energy systems' sustainability and economic feasibility. Identifying typical configurations and the assessment of environomic dependencies is subject to a future publication in redaction.

\newpage

\subsection*{Declaration of generative AI and AI-assisted technologies in the writing process}
During the preparation of this work the authors used \textit{Grammarly} in order to improve the language used in this article. After using this tool, the authors reviewed and edited the content as needed and take full responsibility for the content of the publication.
\subsection*{Data availability}
The datasets for this study can be found in the respective Gitlab repositories:
\begin{itemize}
  \item EnergyScope technologies documentation \newline \url{https://gitlab.com/ipese/energyscope/data/energy-technologies}
  \item LCA technology documentation \newline \url{https://gitlab.com/ipese/energyscope-lca}
\end{itemize}
Additional data is available upon pertinent request.

\subsection*{Acknowledgments}

This  research  has  received  funding  from  (i) the  Flagship Initiative under the project “Flagship PFFS-21-03” \textit{Blue City}.

\subsection*{Competing Interests} The authors declare that they have no
competing financial interests.

\printglossary
\section*{Nomenclature}
The following convention in nomenclature is applied
\begin{itemize}
    \item Modeling variables: $\boldsymbol{\mathrm{X_m^n}}$
    \item Modeling parameters: $\boldsymbol{{X_m^n}}$
    \item Modeling sets: $x \in \mathcal{X-SET}$
    \item General parameters not included in the model: ${{X_m^n}}$
\end{itemize}

\subsection*{Parameters}
\begin{flushleft}
  \begin{tabular}{c l c}
     ${A}$    & Leontief Technology Matrix\\
     ${F}$    & Matrix of elementary flows\\
     $\gls{IW+}$   &  impact assessment leverages the IMPACT World+\\
     ${R}$   & Impact result matrix\\
     ${f}$   & Final demand vector\\
     ${g}$   & \gls{LCI} vector\\
     ${s}$   & Scaling vector\\
     %\\
     $\boldsymbol{c}$    & specific cost      & $\mathrm{[MCHF/\triangle]}$\\
     $\boldsymbol{f_{ext}}$ & Existing capacity    & $\mathrm{[W]}$\\
     $\boldsymbol{lcia}$    & specific \gls{LCIA} impact      & $\mathrm{[LCIA unit/\triangle]}$\\
     $\boldsymbol{n}$    & number      & $\mathrm{[-]}$\\
     $\boldsymbol{t}$    & time      & $\mathrm{[h]}$\\
     \\
     $\boldsymbol{\eta}$   & Efficiency        & $\mathrm{[\%]}$\\
     $\boldsymbol{\tau}$   & Annualisation factor   & $\mathrm{[year^{-1}]}$\\
  \end{tabular}
\end{flushleft}

\clearpage
\subsection*{Variables}
\begin{flushleft}
  \begin{tabular}{c l c}
     $\boldsymbol{\mathrm{C}}$    & Cost           & $\mathrm{[MCHF]}$\\
     $\boldsymbol{\mathrm{F}}$     & Installation size  & $\mathrm{[GW]}$\\
     $\boldsymbol{\mathrm{F_t}}$    & Installation use   & $\mathrm{[\frac{GW\, t_{TP}}{t_{TP}}]}$\\
     $\boldsymbol{\mathrm{LCIA}}$    & \gls{LCIA} Impact           & $\mathrm{[\gls{AoC}-Unit]}$\\
  \end{tabular}
\end{flushleft}

\subsection*{Sets}
\begin{flushleft}
  \begin{tabular}{l l l}
     $\mathcal{COST}$    & Cost  & Investment, Operation and Maintenance\\
     $\mathcal{IND}$      & Indicators   & \\
     $\mathcal{LCIA-I}$    & \gls{LCIA}-Indicators & \glspl{AoC}\\
     $\mathcal{PERIODS}$      & Periods   & \\
     $\mathcal{TEC}$      & Technologies   & \\
  \end{tabular}
\end{flushleft}

\subsection*{Subscripts}
\begin{flushleft}
  \begin{tabular}{l l}
     $inv$      & Investment\\
     $maint$      & Maintenance\\
     $obj$      & Objective\\
     $stat$      & Static\\
     $t$    & Period\\
     $tot$    & Total\\
     $var$      & Variable\\
  \end{tabular}
\end{flushleft}

\newpage

\printbibliography

\newpage
\appendix
\section{Life Cycle Assessment (LCA)}

\subsection{General theory}
The objective of \gls{LCA} is to assess the potential impacts of a product system throughout its whole life cycle. The typical model employed is the cradle-to-grave approach, commencing with raw material extraction and culminating in product disposal via recycling, landfilling, or incineration \cite{international_organization_for_standardization_iso_2006}. The \gls{LCA} procedure encompasses four stages: 
\begin{enumerate}
    \item Definition of goal and scope
    \item \acrlong{LCI} (\acrshort{LCI})
    \item \acrlong{LCIA} (\acrshort{LCIA})
    \item Interpretation
\end{enumerate}

\subsubsection{Definition of Goal and Scope}

Each technology must be delineated distinctly to accurately incorporate \gls{LCA} into the \gls{ES} model. A dual-product specification—of its construction and operation—characterizes every technology, mirroring the cradle-to-grave approach. The functional units are contingent upon the technology's end-use category, as detailed in Table \ref{tab:unit}.

\subsubsection{Life Cycle Inventory (LCI)}

Instead of a sequential method, the \gls{LCI} computation employed a matrix approach. This method facilitates the seamless integration of feedback loops in product systems. It negates the need for explicit scaling factor computation, provided the final demand vector $f$ is known (Table \ref{tab:lca_convention}). The relationships between products are described by the technology matrix $A$. Further matrix structure and relationship specifications are available within the ecoinvent database \cite{wernet_ecoinvent_2016}. The LCI's expressions in terms of matrices are as follows:
\begin{align}
    s &= A^{-1} \cdot f\\
    g &= B \cdot s = B \cdot \left( A^{-1} \cdot f \right )
\end{align}

\subsubsection{Life Cycle Impact Assessment}

The \gls{LCIA} uses twenty-nine damage level indicators, which are further complemented by three \glspl{AoC} (\gls{CF}, \gls{FNEU} and \gls{WSF}) and two \glspl{AoP} (\gls{REQD}, \gls{RHHD}). The impact assessment leverages the \gls{IW+} method \cite{bulle_impact_2019},  which matrix is multiplied by the LCI's vector $g$ to get the impact result vector $h$:
\begin{equation}
    h = \mathrm{C} \cdot g
    \label{eq:R}
\end{equation}

\begin{table}[!htb]
\centering
\caption[\gls{LCA} notation]{Notation used in the life cycle inventory phase of the \gls{LCA}}
\begin{tabular}{@{}cc@{}}
\toprule
\textbf{Notation} & \textbf{Name}     \\ \midrule
A        & Leontief technology matrix  \\
B        & Matrix of elementary flows \\
$f$      & Final demand vector               \\
$s$        & Scaling vector             \\
$g$      & Life cycle inventory vector       \\ \bottomrule
\end{tabular}
\label{tab:lca_convention}
\end{table}

\subsubsection{Assumptions}

In mobility, car products from the ecoinvent database are specified per kilometer. For Switzerland, an average of 1.6 passengers per car was assumed, as cited in \cite{schnidrigModellingFrameworkAssessing2021}. Although impacts from technology maintenance and transportation are predominantly encapsulated within the operation, the granularity of the details might vary. Consequently, the decision was to integrate these impacts into the operational facet rather than treating them separately.
\clearpage
\subsection{Functional units}
\begin{landscape}
    \centering
    \tiny
\begin{longtable}{p{4cm}p{2cm}p{4cm}p{3cm}p{1cm}p{1.5cm}p{1.5cm}p{2cm}}
    \caption{\gls{ES}-Technology \gls{LCIA}-Activity mapping with type region, conversion factor and conversion factor unit.}\\
    \label{tab:app_tech_map}\\
        \toprule
        Technology(EnergyScope) & Type & Name of the activity (LCA) & Product of the activity (LCA) & Region (LCA) & Source & Conversion factor & Unit \\ \midrule
        \endfirsthead
        \toprule
        Technology (EnergyScope) & Type & Name of the activity (LCA) & Product of the activity (LCA) & Region (LCA) & Source & Conversion factor & Unit \\ \midrule
        \endhead
        Alkaline Electrolysis & Construction & electrolyzer production, 1MWe, AEC, Stack & electrolyzer, 1MWe, AEC, Stack & RER & Sacchi et al. & 1492.537313 & unit / GW \\ 
        Alkaline Electrolysis & Operation & hydrogen production, gaseous, 20 bar, from AEC electrolysis, from grid electricity & hydrogen, gaseous, 20 bar & CH & Sacchi et al. & 30030.03003 & kg / GWh \\ 
        Aluminum Making & Construction & aluminium casting facility construction & aluminium casting facility & RER & ecoinvent 3.8 cut-off & 67.38461538 & unit / kt/h \\ 
        Aluminum Making & Operation & aluminium production, primary, ingot & aluminium, primary, ingot & IAI Area, EU27 \& EFTA & ecoinvent 3.8 cut-off & 1000000.0 & kg / kt \\ 
        Aluminum Making Heat Recovery & Construction & aluminium casting facility construction & aluminium casting facility & RER & ecoinvent 3.8 cut-off & 67.38461538 & unit / kt/h \\ 
        Aluminum Making Heat Recovery & Operation & aluminium production, primary, ingot & aluminium, primary, ingot & IAI Area, EU27 \& EFTA & ecoinvent 3.8 cut-off & 1000000.0 & kg / kt \\ 
        Anaerobic Digestion of Sewage Sludge & Construction & anaerobic digestion plant construction, for biowaste & anaerobic digestion plant, for biowaste & CH & ecoinvent 3.8 cut-off & 63.02158273 & unit / GW \\ 
        Anaerobic Digestion of Sewage Sludge & Operation & anaerobic digestion of manure & biogas & RoW & ecoinvent 3.8 cut-off & 100478.2766 & m3 / GWh \\
\bottomrule
    \end{longtable}
\end{landscape}

\clearpage
\subsection{EnergyScope technologies LCI database}
\begin{figure}[htb]
\centering
  \begin{subfigure}[b]{0.45\textwidth}
         \centering
	\includegraphics[width=1\textwidth]{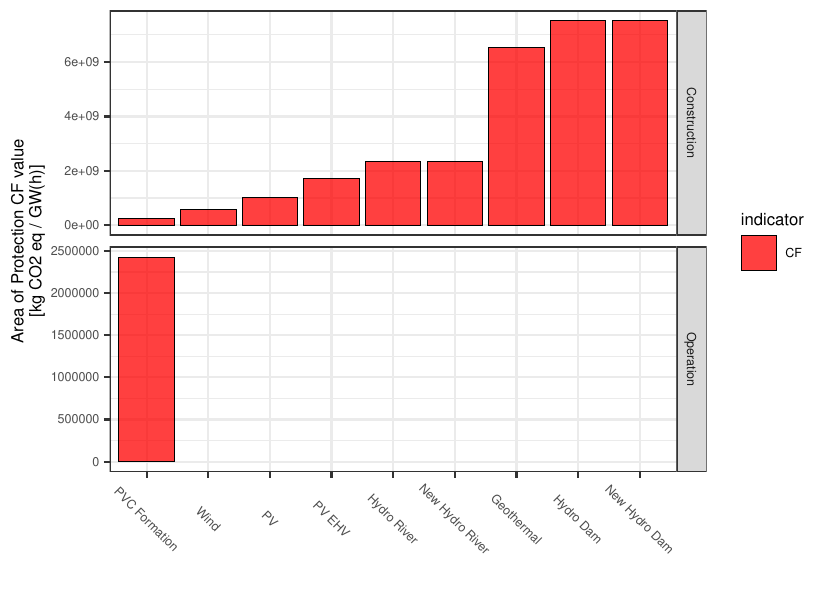}
    \caption{Carbon Footprint}
    \label{fig:LCA_DB_CF}
    \end{subfigure}
   \begin{subfigure}[b]{0.45\textwidth}
         \centering
         \includegraphics[width=1\textwidth]{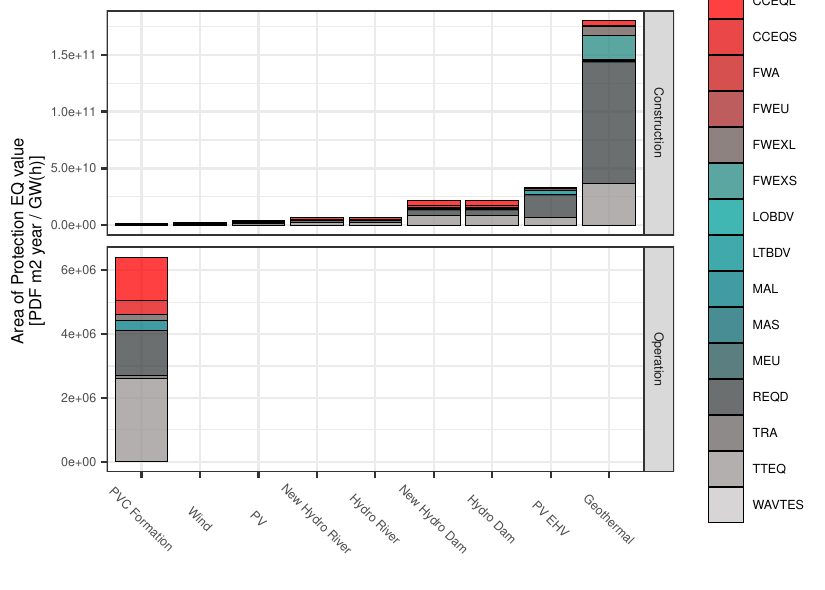}
    \caption{Ecosystem Quality}
    \label{fig:LCA_DB_EQ}
    \end{subfigure}
    
   \begin{subfigure}[b]{0.45\textwidth}
         \centering
         \includegraphics[width=1\textwidth]{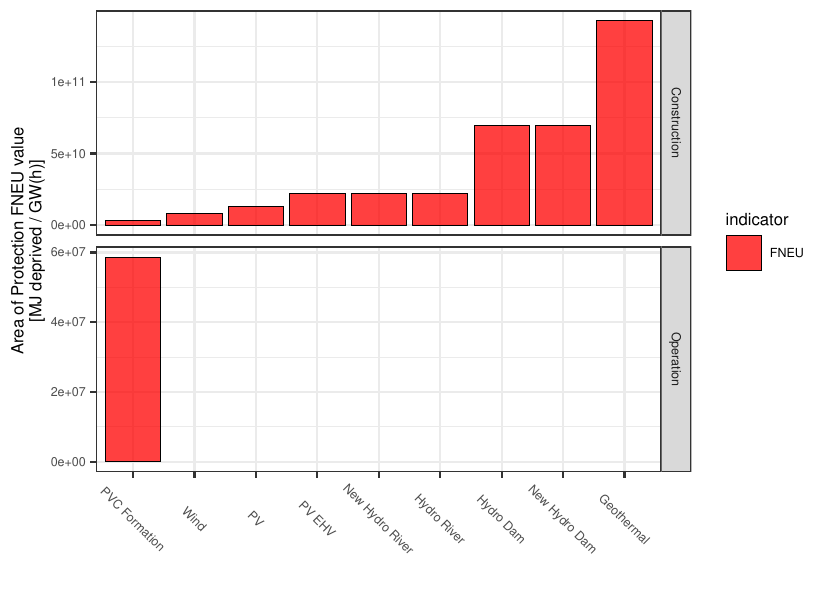}
    \caption{Fossil \& Nuclear Energy Use}
    \label{fig:LCA_DB_FNEU}
    \end{subfigure}
    \begin{subfigure}[b]{0.45\textwidth}
         \centering
         \includegraphics[width=1\textwidth]{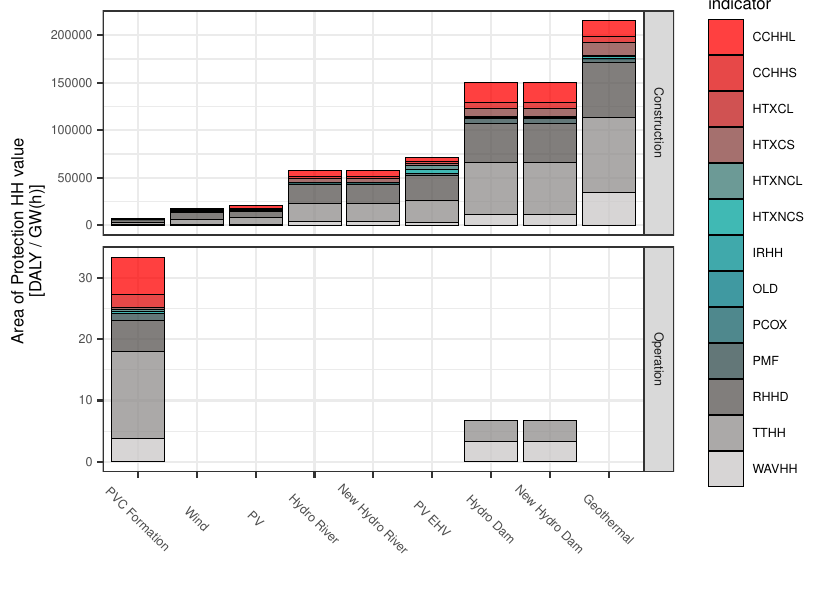}
    \caption{Human Health}
    \label{fig:LCA_DB_HH}
    \end{subfigure}
    
   \begin{subfigure}[b]{0.45\textwidth}
         \centering
         \includegraphics[width=1\textwidth]{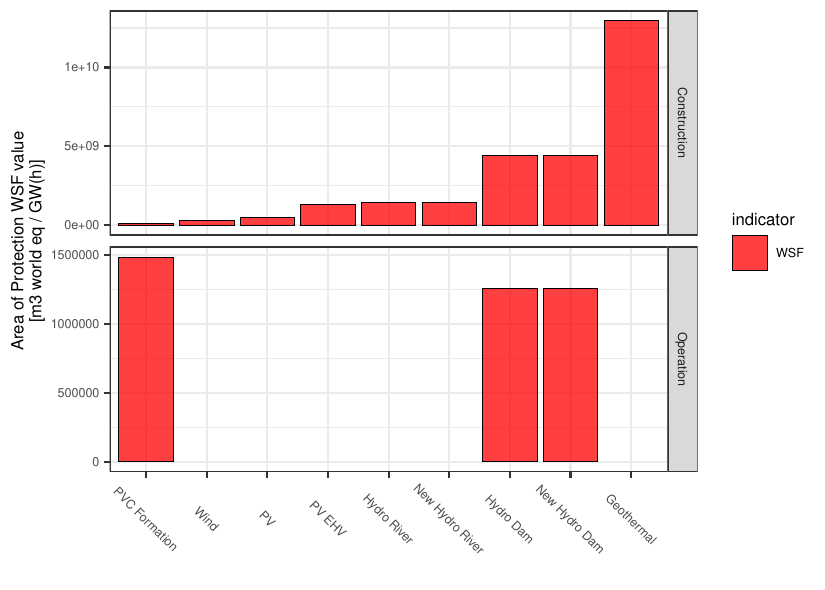}
    \caption{Water Scarcity Footprint}
    \label{fig:LCA_DB_WSF}
    \end{subfigure}   \caption{LCIA impact technologies characterization for the renewable energy sector.}
    \label{fig:LCA_db}
\end{figure}

A holistic database for all technologies in EnergyScope has been created, allowing to assess the environmental impact of all technologies throughout the End-Points, Areas of Protection and Areas of Concern. Comparison in between technologies of one sector can directly be proceeded, furthermore allowing to identify which impact weighs most in the Areas of Concern for example. Representative for all technologies, the renewable energy sector technologies is presented in Figure \ref{fig:LCA_db}. The entire database can be found \url{https://gitlab.com/ipese/energyscope-lca/Data/LCA-technologies}.

\clearpage

\subsection{Double-counting removal}

\begin{algorithm}[!hp]
\caption{is\_market(A, DB)}\label{alg:double_counting_markets}
\begin{algorithmic}[1]
\State{\textbf{Input}: activity A, database DB}
\State{\textbf{Output}: list L of activities to perform double-counting removal on}
\State{condition} \Comment{condition to know whether the activity is a market or not}
\If{condition}
    \State{save(A, DB)}
    \For{parent in technosphere\_flows(A):}
        \State{is\_market(parent, DB)}
    \EndFor
\Else
    \State{L.append(A)}
\EndIf 
\end{algorithmic}
\end{algorithm}

%%%%%%%%%%%%%%%%%%%%%%%%%%%%%%%%%%%%%%%%%%%%%%%%%%%%
%%%%%%%%%%%%%%%%%%%%%%%%%%%%%%%%%%%%%%%%%%%%%%%%%%%%
%%%%%%%%%%%%%%%%%%%%%%%%%%%%%%%%%%%%%%%%%%%%%%%%%%%%
%%%%%%%%%%%%%%%%%%%%%%%%%%%%%%%%%%%%%%%%%%%%%%%%%%%%
%%%%%%%%%%%%%%%%%%%%%%%%%%%%%%%%%%%%%%%%%%%%%%%%%%%%

\section{Technology distributions}
\label{app:SOO}
\subsection{Energy system configurations}
\label{app:SOO_sectoral_cost}
\begin{figure}[htb]
\centering
  \begin{subfigure}[b]{0.3\textwidth}
         \centering
      \includegraphics[width=1\textwidth]{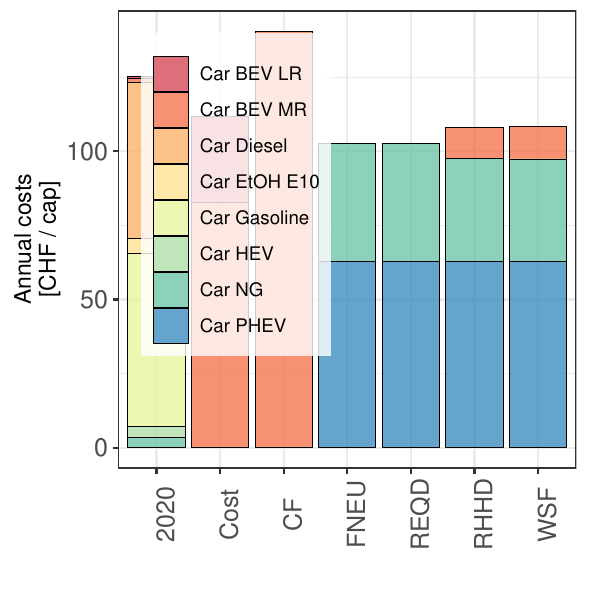}
    \caption{Mobility Passenger Private}
    \label{fig:cap_mob_pri}
    \end{subfigure}
   \begin{subfigure}[b]{0.3\textwidth}
         \centering
      \includegraphics[width=1\textwidth]{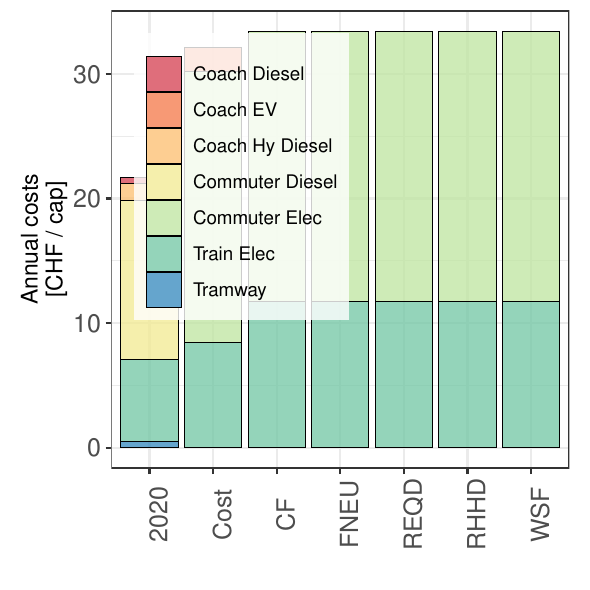}
    \caption{Mobility Passenger Public}
    \label{fig:cap_mob_pub}
    \end{subfigure}
   \begin{subfigure}[b]{0.3\textwidth}
         \centering
      \includegraphics[width=1\textwidth]{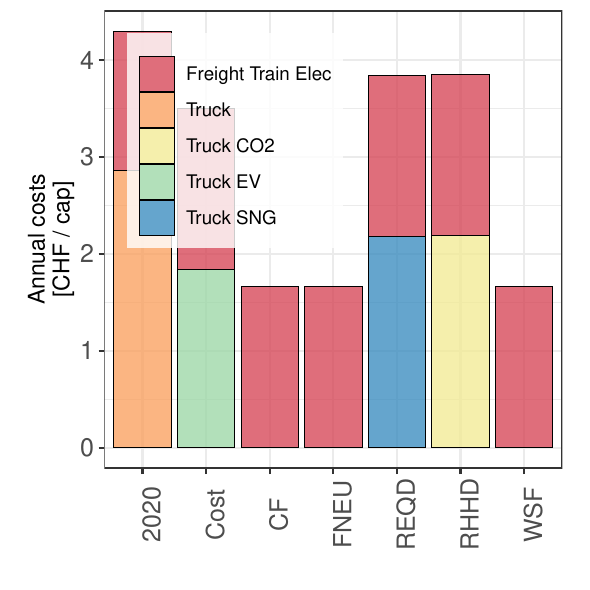}
    \caption{Mobility Freight}
    \label{fig:cap_mob_freight}
    \end{subfigure}
    
    \begin{subfigure}[b]{0.3\textwidth}
         \centering
      \includegraphics[width=1\textwidth]{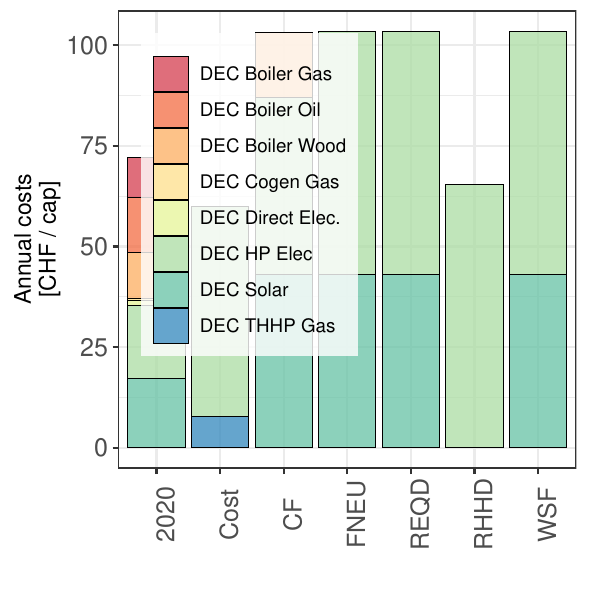}
    \caption{Heating Decentralized}
    \label{fig:cap_lt_dec}
    \end{subfigure}
   \begin{subfigure}[b]{0.3\textwidth}
         \centering
      \includegraphics[width=1\textwidth]{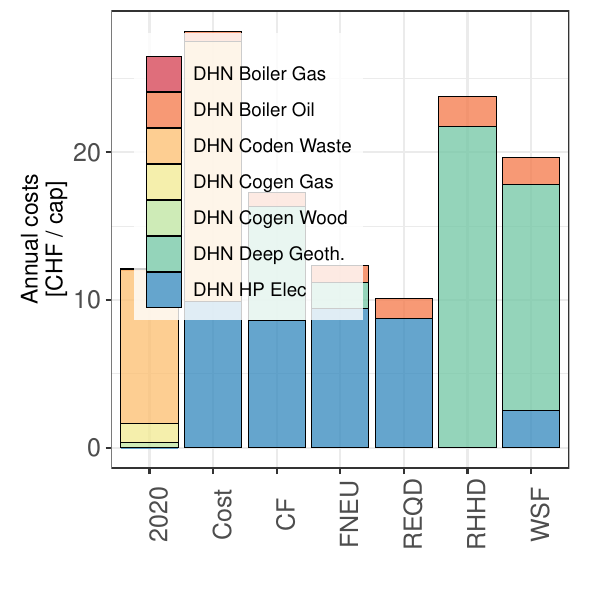}
    \caption{Heating DHN}
    \label{fig:cap_lt_dhn}
    \end{subfigure}
   \begin{subfigure}[b]{0.3\textwidth}
         \centering
      \includegraphics[width=1\textwidth]{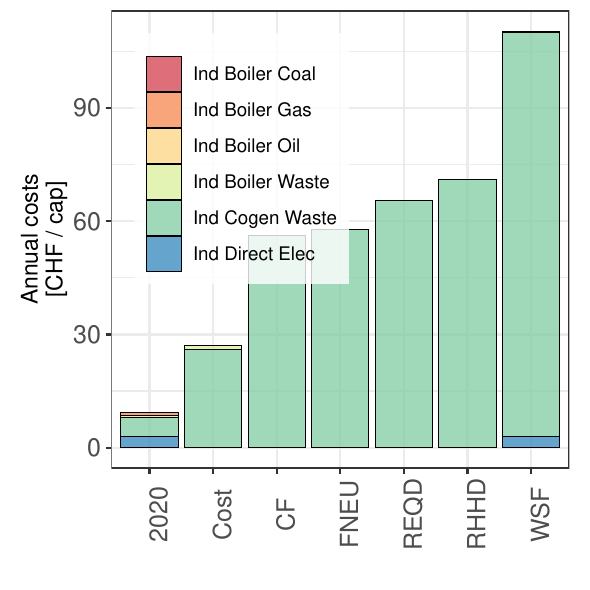}
    \caption{Heating Industry}
    \label{fig:cap_ht}
    \end{subfigure}
   \caption{Energy system cost compositions of the reference and environomic optimization scenarios for the mobility and heating sector.}
    \label{fig:cap_sub}
\end{figure}

The shift from fossil fuel imports to local energy resources leads to a shift in the \textbf{mobility} sector too, where the fossil fuel dominated private and freight transportation (Figures \ref{fig:cap_mob_pri}, \ref{fig:cap_mob_pub} \& \ref{fig:cap_mob_freight}) is replaced by pure battery electric vehicles for the cost and \gls{CF} scenarios and a vehicle fleet based on synthetic fuels composed of hybrid electric vehicles and pure \ch{CH4} vehicles for the remaining environmental \gls{OF} minimizations. Electric public transportation is maximized, similar to electric freight transport on rails. While \gls{CF}, \gls{FNEU}, and \gls{WSF} minimizations all select Hydrogen Fuel Cell trucks for road transportation, other truck types are selected for the remaining scenarios (cost: battery electric trucks, \gls{REQD}: \ch{CH4} powered trucks \& \gls{RHHD}: the carbon capture trucks fueled by synthetic Diesel).

The same behavior can be observed in the \textbf{heating} sector, where fossil fuels are replaced by electric, more efficient, technologies maximizing the use of district solutions. While at decentralized level (Figure \ref{fig:cap_lt_dec}) the heating technologies are dominated by heat pumps complemented by solar thermal panels (except \gls{RHHD}) in combination with Gas powered heat pumps in the cost minimization and wood boilers in \gls{CF} minimization, the district heating (Figure \ref{fig:cap_lt_dhn}) configurations differ more: \gls{CF} and \gls{FNEU} are based on district scale heat pumps, \gls{RHHD} and \gls{WSF} on geothermal heating and cost combines heat pumps with waste cogeneration. While these district solutions are based on baseline technologies and thus subject to intermittency, backup solutions at district levels are installed based on synthetic fuels to back up the peak heating demands. Throughout the scenarios process heat for industry (Figure \ref{fig:cap_ht}) is generated on waste-based cogeneration plants, where \gls{WSF} minimization uses the synergy with the district heating to provide additional low-temperature heat.

The use of local resources only leads to energy \textbf{storage} demands (Figure \ref{fig:capex_overview} secondary y-axis). Throughout the scenarios, hydro dam storage is used at its maximum capacity of \SI{8.9}{TWh}. The use of \ch{CH4} as heating peak backup and the balancing of the dephasing between seasonal energy demand and production is balanced in the cost configuration with the installation of \SI{2}{TWh} of biogenic \ch{CH4} storage. The production of synthetic fuels for mobility based on electrolysis and methanation requiring \ch{CO2} storage can be observed in all environmental minimization scenarios except \gls{CF}, where the fleet is electric. An additional energy storage capacity in the form of \SI{3.5}{TWh} of Diesel is required in the \gls{RHHD} scenario, where the Carbon Capture truck is used for road freight mobility.

\subsection{Sectoral environmental impacts}
\label{app:SOO_sectoral_env}
\begin{figure}[!htb]
	\centering
	\includegraphics[width=\textwidth]{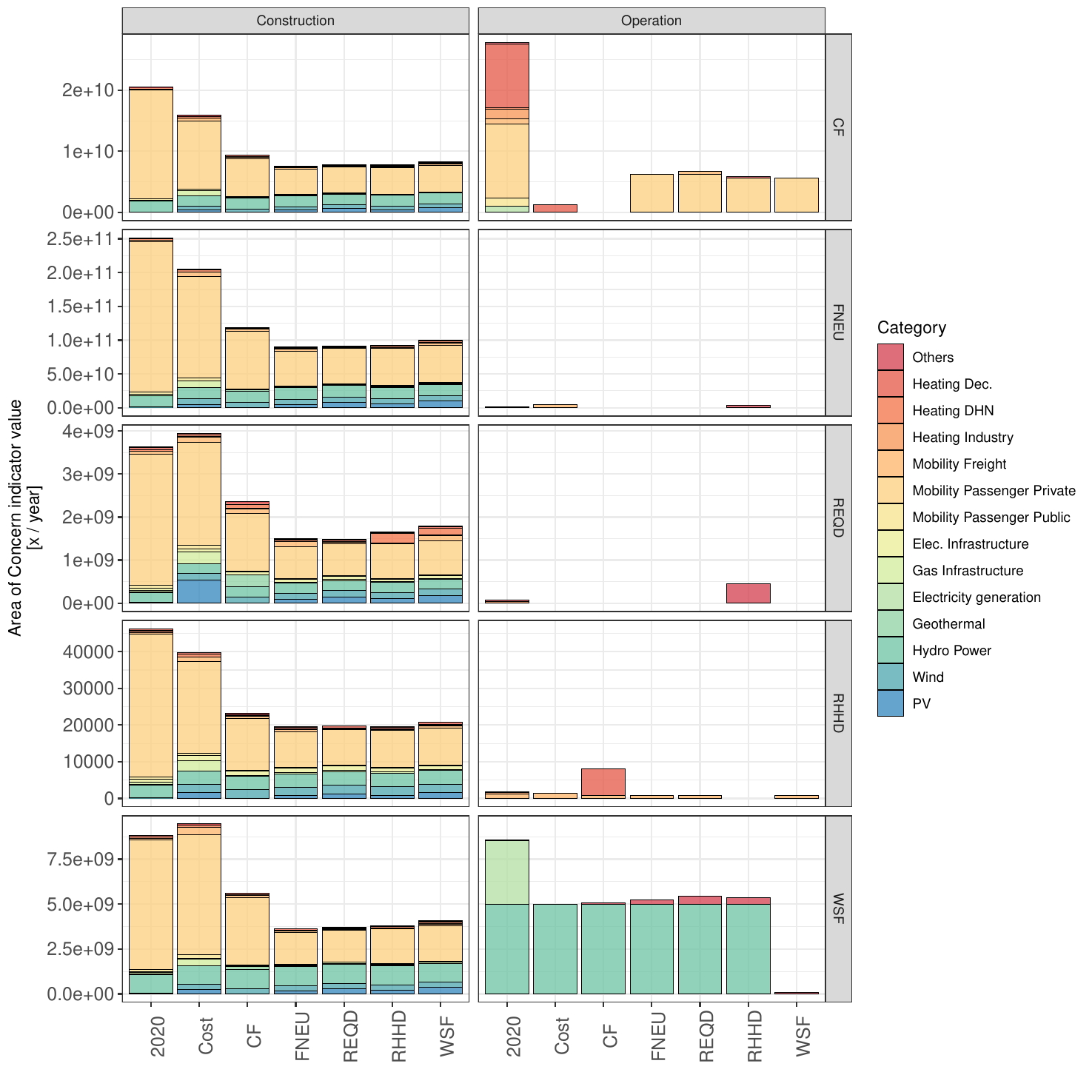}
	\caption{LCIA impacts composition for \gls{SOO}. Case study Switzerland 2020 independent, without nuclear power.}
	\label{fig:lcia_overview}
\end{figure}

By analyzing the relative contribution to overall LCA and cost impact scores in term of construction  (related to capital infrastructure) and operation (variable energy resource supply and cost) as depicted in Figure \ref{fig:lcia_overview} where the first column corresponds to the 2020 baseline scenario and the remaining columns represent the histograms of the single-score optimizations, we can observe a shift from operational to capital cost and impacts. While the current energy system predominantly reflects operational environmental impacts, the LCIA indicators for economic configurations lean more towards constructional impacts. This inclination suggests a strategic minimization of specific technology constructions and uses at the expense of other indicators.

Examining the constant emissions (construction) in the left column of Figure \ref{fig:lcia_overview}, the private mobility sector is the major contributor accounting for between \SI{93}{\%} and \SI{46}{\%} of the impact across all \glspl{AoC}. The residual impact is attributed to renewable energy. Here, hydropower plants have the highest impacts, trailed by \gls{PV}, with the peak observed in the cost minimization scenario. Notably, the composition of these impacts remains consistent across different impact categories.

When scrutinizing operational impacts, all optimized scenarios display reduced impact compared to 2020, as the system relies on local, more efficient renewable resources. Operational impacts for indicators \gls{FNEU}, \gls{REQD}, and \gls{RHHD} are minimal. In contrast, \gls{CF} and \gls{WSF} demonstrate impacts comparable to constructional ones unless these objectives are minimized. Specifically, when \gls{CF} is minimized, the direct emissions are virtually eradicated, thanks to the adoption of carbon-neutral mobility and heating technologies. The water scarcity footprint, predominantly influenced by hydro dam usage and 2020's nuclear plants, is reduced by sidelining hydro dams and compensating the energy shortfall with additional PV installations.

Transitioning from the current energy system to an environmentally optimized one inherently diminishes the indirect impact across all indicators and the direct carbon emissions. While direct impacts for \gls{FNEU}, \gls{REQD}, and \gls{RHHD} remain minimal in both the current and optimized systems relative to their indirect counterparts, \gls{WSF} at the operational level are heavily influenced by existing hydro dams. When minimizing its impact, this creates a paradoxical situation, leading to underutilizing existing infrastructure.

\paragraph{From variable to constant emissions}
%From direct to indirect emissions
The study categorizes \gls{LCA} indicators into constant emissions from construction and variable emissions from operation. Current energy systems show higher operational impacts, while economically optimized configurations skew towards construction impacts, suggesting a trade-off in technology use. The private mobility sector dominates construction impacts, ranging from \SI{93}{\%} to \SI{46}{\%}. Hydropower displays the most considerable construction emissions, followed by photovoltaics, especially in the cost-minimization scenario.

In operation, optimized scenarios reduce emissions across all indicators compared to 2020. Direct emissions for \gls{FNEU}, \gls{REQD}, and \gls{RHHD} are minimal, but \gls{CF} and \gls{WSF} operational impacts can rival construction emissions unless they are targeted for minimization. \gls{CF} minimization almost eliminates its operational emissions through carbon-neutral mobility and heating. \gls{WSF} impact lessens when hydro dams are underused in favor of additional \gls{PV} installations to address energy deficits.

Transitioning to optimized energy systems significantly lowers indirect and direct carbon emissions across all indicators. Direct impacts for \gls{FNEU}, \gls{REQD}, and \gls{RHHD} are minor relative to their indirect counterparts in both current and optimized systems. However, \gls{WSF} operational impact is significantly affected by hydro dams, presenting an economic challenge when trying to minimize its impact due to the potential underutilization of this infrastructure.

\end{document}